\documentclass[aps,prx,reprint,floatfix,showpacs,groupaddress]{revtex4-2}

\usepackage{amssymb}
\usepackage{amsmath}
\usepackage{bm}
\usepackage{graphicx}
\usepackage{hyperref}
\DeclareMathOperator{\Tr}{Tr}
\usepackage[utf8]{inputenc}
\newcommand{\Q}{\mathbf{Q}}
\newcommand{\n}{\mathbf{\hat{n}}}
\newcommand{\OMEGA}{\bm{\hat{\Omega}}}
\newcommand{\T}{\mathbf{\hat{T}}}
\newcommand{\q}{\mathbf{\hat{q}}}
\newcommand{\w}{\mathbf{\hat{w}}}
\newcommand{\tphi}{\tilde{\varphi}}
\newcommand{\tn}{\mathbf{\tilde{n}}}
\newcommand{\XI}{\bm{\hat{\Xi}}}

\begin{document}
\title{Kinematics and dynamics of disclination lines in three-dimensional nematics}
\author{Cody D. Schimming}
\email{cschim@lanl.gov}
\affiliation{Theoretical Division and Center for Nonlinear Studies (CNLS), Los Alamos National Laboratory, Los Alamos, New Mexico 87545, USA}

\author{Jorge Vi\~nals}
\affiliation{School of Physics and Astronomy, University of Minnesota, Minneapolis, Minnesota 55455, USA}

\begin{abstract}
    An exact kinematic law for the motion of disclination lines in nematic liquid crystals as a function of the tensor order parameter $\Q$ is derived. Unlike other order parameter fields that become singular at their respective defect cores, the tensor order parameter remains regular. Following earlier experimental and theoretical work, the disclination core is defined to be the line where the uniaxial and biaxial order parameters are equal, or equivalently, where the two largest eigenvalues of $\Q$ cross. This allows an exact expression relating the velocity of the line to spatial and temporal derivatives of $\Q$ on the line, to be specified by a dynamical model for the evolution of the nematic. By introducing a linear core approximation for $\Q$, analytical results are given for several prototypical configurations, including line interactions and motion, loop annihilation, and the response to external fields and shear flows. Behaviour that follows from topological constraints or defect geometry is highlighted. The analytic results are shown to be in agreement with three dimensional numerical calculations based on a singular Maier-Saupe free energy that allows for anisotropic elasticity. 
\end{abstract}

\maketitle

\section{Introduction}\label{sec:Intro}

Topological defects play an integral role in the response and nonequilibrium evolution of many physical systems: in type-II superconductors, for example, vortices allow magnetic field lines to penetrate the material and dissipate \cite{abrikosov57,chaikin95}; in solids, dislocations mediate plastic deformation and melting \cite{halperin78,nelson79,young79,kleman08}; in developing biological tissue, defects indicate sites of further morphogenesis and curvature generation \cite{etournay16,livshits17,hoffman22}; and in nematic liquid crystals, disclinations promote aggregation of colloidal particles and generate fluid velocity in active materials \cite{ravnik07,copar11,doo18,opathalage19,duclos20}. Therefore significant efforts are under way to further elucidate the general principles behind their dynamics \cite{peach50,blatter94,chaikin95,pismen99,olson01,svensek02,kremen16,tang17,audun18,acharya20,binysh20,beliaev21,long21,houston21}. A common theoretical strategy is to treat defects as effective \lq\lq particles.'' This is made possible by topological constrains: defects cannot spontaneously disappear or nucleate, but must instead pairwise annihilate or unbind (similar to particles and antiparticles). Additionally, topological defect charges can be quantized, that is, related to the non-trivial homotopy group of the physical system \cite{pismen99,alexander12}. Thus, in those cases in which the response or the temporal evolution of a system are determined by the nature and distribution of defects (with the overall dynamics slaved to such a distribution), one needs only focus on laws of motion for the effective ``particles'' (or ``strings'' or ``membranes'' for higher dimensional defects).

Our focus here is on topological defects in nematic phases (disclinations). In a nematic, the order parameter is a symmetric, traceless tensor, $\Q$, which captures both rotational and apolar symmetries. Nematics are well known for their anisotropic optical and hydrodynamic properties \cite{deGennes75,beris94,yeh09}; however, there has been increasing interest in the role of disclinations in both passive and active nematics. In the former, disclinations mediate colloidal aggregation and can be patterned to engineer transport throughout the material \cite{gu00,ravnik07,copar11,re:peng15,baba18,re:turiv20,guo21}. In the latter, disclinations form spontaneously, and generate flows depending on their topological or geometric character \cite{doo16b,opathalage19,duclos20,binysh20}.

In this work, a particle-field transformation is introduced to describe disclination line motion in nematic phases. Such a transformation relating the location of the line to the field equations governing the evolution of the nematic tensor order parameter allows for an exact kinematic law of motion for the disclination, independent of the microscopic model governing the evolution of nematic order. The type of particle-field transformation that we introduce has been successfully used to analyze and track vortex motion in superfluids \cite{halperin81}, the motion of point defects in $n$-vector models \cite{liu92,mazenko97,mazenko99}, and, more recently, the motion of dislocations in solids in both two \cite{audun18,re:skaugen18b} and three dimensions \cite{skogvoll22}. The method has also been used to describe disclination motion in nematic active matter, albeit in two spatial dimensions \cite{angheluta21}.

Despite recent interest in the motion of disclinations in nematic phases, there are still many open questions regarding their structure and motion, particularly in three dimensions. While the topological structure is deceptively simple (the first fundamental group in three dimensions is $\mathbb{Z}_{2}$ instead of $\mathbb{Z}$ in two dimensions), the geometric character of the defect is completely different. Nematic disclinations in two dimensions are point defects, whereas line disclinations in three dimensions are spatially extended, and generally described by two independent vectors: the tangent vector to the disclination $\T$, and the rotation vector describing the nematic distortion near the defect $\OMEGA$ \cite{friedel69,deGennes75}. Further, and unlike all the applications of particle-field transformations mentioned above, the tensor order parameter is regular at the disclination core \cite{schopohl87,schimming20,schimming21}. Hence the core is not defined by the well studied director phase singularity, rather by a crossing of eigenvalues of the tensor order parameter. 

We first introduce the particle-field transformation to a nematic phase in three dimensions described by a tensor order parameter $\Q$. Even though the order parameter does not go to zero at defect locations, a quantity defined on a subspace of the order parameter space does, a fact that is used to locate disclination cores. The transformation leads to a kinematic law which is then used to obtain analytic predictions of disclination motion using suitable approximations of the order parameter in the vicinity of the core. Disclination velocity is seen to be determined by gradients of the tensor order parameter at the core, a fact that significantly simplifies consideration of a number of prototypical configurations involving lines, loops, and their interactions. Within a linear core approximation for the order parameter, we obtain analytic results for the evolution of both optimally oriented and twisted defect configurations, including elastic anisotropy, find transverse defect mobilities of topological origin, study disclination line interaction and recombination, loop shrinkage, and show that external fields or imposed shear flows can spatially separate (sort) lines and loops according to their topological charge distribution. Transverse mobilities and defect sorting are important for the many applications that rely on defect control and defect engineering which are currently under development in soft, active, and biological matter. Importantly, the motion that follows derives not only from the forces among disclination segments, but it also incorporates the necessary topological constraints explicit in the kinematic law. Although our main focus is on nematic liquid crystals, the techniques described should apply to a range of complex systems in which topological defects are allowed.

The rest of the paper is organized as follows: In Sec. \ref{subsec:CoreStructure} we briefly review the structure of the disclination core in three-dimensional nematics. In Sec. \ref{sec:LineKinematics} we use the fact that the eigenvalues of the order parameter $\Q$ cross at the disclination core to derive a kinematic velocity equation in terms of the order parameter. In Sec. \ref{sec:Approximations} we describe an analytical approximation of $\Q$ near the core (the \lq\lq linear core approximation"), and show how this approximation may be used to obtain the velocity of a disclination in the presence of an imposed rotation of the director (this is the analog of the Peach-Koehler force in elasticity theory). In Sec. \ref{sec:Results} we present analytical predictions for the motion of disclinations in both two-dimensional and three-dimensional configurations involving disclination annihilation and exposure to external fields and flows. Throughout we supplement our analysis with numerical calculations in two- and three-dimensions, and find excellent agreement between the two despite the complicated nonlinearities present in the computational model. Finally, in Sec. \ref{sec:Discussion} we discuss our results and their implications for nematics and other systems in which disclinations are pervasive. We also discuss potential further work in understanding disclination dynamics.

\section{Disclination core structure}\label{subsec:CoreStructure}

Consider an ensemble of nematogens, each described by a unit vector $\hat{\bm{\xi}}$ giving its molecular orientation. Let $p(\hat{\bm{\xi}})$ be the equilibrium probability density of orientations at constant temperature, defined on the unit sphere ${\cal S}^{2}$. The tensor order parameter is defined as 
\begin{equation} \label{eqn:QDef}
    \Q = \int_{{\cal S}^2}\left(\bm{\hat{\xi}}\otimes\bm{\hat{\xi}} - \frac{1}{3}\mathbf{I}\right)p(\bm{\hat{\xi}})\, d\Sigma(\bm{\hat{\xi}})
\end{equation}
where $\mathbf{Q}$ can be uniform or a field if distorted configurations at the mesoscale are considered. With this definition, the tensor $\mathbf{Q}$ is symmetric and traceless, and can be represented as
\begin{equation} \label{eqn:QParam}
    \Q = S\left[\n\otimes\n - \frac{1}{3}\mathbf{I}\right] + P\left[\mathbf{\hat{m}} \otimes \mathbf{\hat{m}} - \bm{\hat{\ell}}\otimes \bm{\hat{\ell}}\right]
\end{equation}
where $S$ and $P$ are the uniaxial and biaxial order parameters respectively, $\hat{\mathbf{n}}$ is the uniaxial director, $ \{\n,\mathbf{\hat{m}},\bm{\hat{\ell}}\}$ form an orthonormal triad, and $\mathbf{I}$ is the $3\times3$ unit matrix. The eigenvectors of $\Q$ give the orientation of the nematic, i.e. the director $\n$, while the eigenvalues of $\Q$ represent the degree of ordering in the nematic. For a uniaxial nematic, $\Q$ is simply $\Q = S\left[\n \otimes \n - (1/3)\mathbf{I}\right]$. The scalar $S$ gives the local degree of ordering. $S = 0$ indicates the system is in the isotropic phase, while $S > 0$ indicates the system is in the nematic phase.

\begin{figure}
    \centering
    \includegraphics[width = \columnwidth]{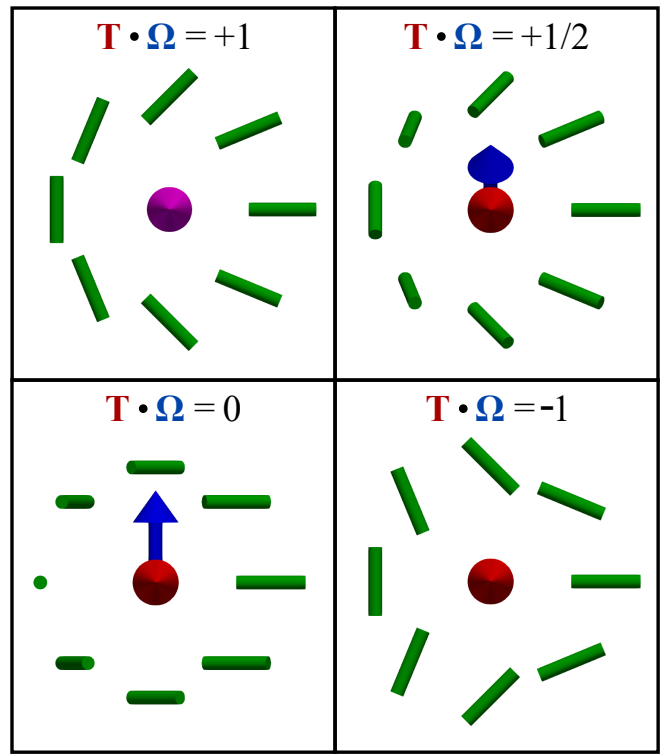}
    \caption{Examples of director configurations around disclinations with varying $\T \cdot \OMEGA$. The panels indicate that a ``$+1/2$ wedge'' type disclination may be continuously rotated into a ``$-1/2$ wedge'' disclination in three dimensions. While these two configurations are topologically distinct in two dimensions, they are topologically equivalent in three dimensions.}
    \label{fig:3DDisclinations}
\end{figure}

Macroscopically, a disclination line is a continuous line in which the director, $\n$, is singular. Its geometry is characterized by its local tangent vector $\T$ and a rotation vector $\OMEGA$.
Near the singular core, $\OMEGA \cdot \n = 0$ on the plane normal to $\T$ \cite{friedel69}. That is, close to the disclination core, the director lies on a plane perpendicular to $\OMEGA$ as one encircles the core on its normal plane. Both $\T$ and $\OMEGA$ may vary along the disclination, and their relationship (i.e. $\T \cdot \OMEGA$) greatly affects its local motion \cite{duclos20,binysh20,long21}. Figure \ref{fig:3DDisclinations} illustrates director configurations on the plane normal to the line for various values of $\T \cdot \OMEGA$. A few special cases referenced throughout the paper include $\T \cdot \OMEGA = +1$ (a $+1/2$ wedge disclination), $\T \cdot \OMEGA = -1$ (a $-1/2$ wedge disclination), and $\T \cdot \OMEGA = 0$ (a twist disclination). The wedge disclination nomenclature follows from their analog in two dimensions, thus carrying over the $+1/2$ and $-1/2$ charge in 2D, while the twist type disclination is named because the twist elastic distortion is the only elastic distortion present in the configuration. Even though these cases are geometrically distinct, they are all topologically equivalent (the topological charge of a disclination line in three dimensions is always $1/2$). A general disclination line has $\T \cdot \OMEGA$ varying along its contour. This is quite different from a dislocation in a solid, in which the topological invariant is the Burgers vector $\mathbf{b}$, constant along the line. While useful analogies have been made between $\OMEGA$ and $\mathbf{b}$ \cite{long21}, they are mathematically distinct, as the Burgers vector is topologically protected, and the rotation vector is not.

\begin{figure}
    \centering
    \includegraphics[width = \columnwidth]{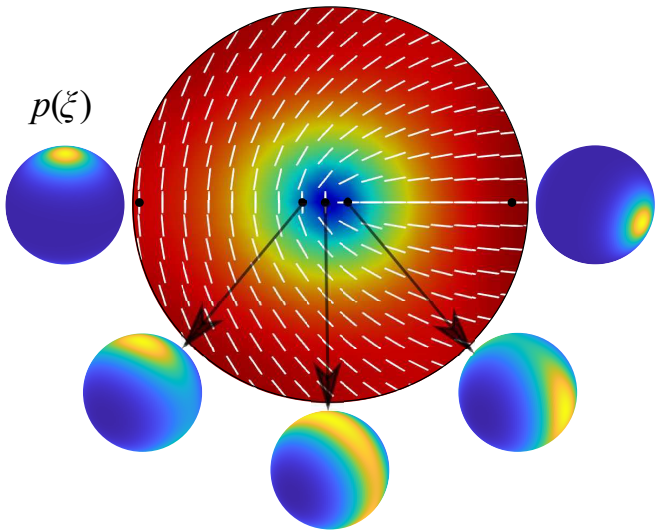}
    \caption{Computed nematogen orientational distribution, $p(\bm{\hat{\xi}})$ (Eq. \eqref{eqn:QDef}) at various points through a wedge disclination (color indicates uniaxial order $S$, while white lines show director $\n$). Far from the disclination core the distribution is uniaxial. As the core is approached the distribution becomes biaxial. At the core the distribution is again uniaxial such that nematogens are all equally likely to lie in the plane perpendicular to $\OMEGA$. The distribution has been computed by the method of singular potentials as outlined in \cite{schimming21}.}
    \label{fig:Distribution}
\end{figure}

More microscopically, the structure of a disclination is illustrated in Fig. \ref{fig:Distribution} which shows the probability distribution on the unit sphere at various locations in its vicinity (see Sec. \ref{subsec:Implementation} and Appendix \ref{append:CompDetails} for further details). Far from the disclination, the distribution is uniaxial (fluctuations from the primary direction are isotropically distributed, and $\Q$ has two degenerate eigenvalues). As the core is approached, the distribution spreads out in the direction perpendicular to $\OMEGA$, becoming biaxial, so that the order parameter $\Q$ has three distinct eigenvalues ($P > 0$). Exactly at the core, the distribution becomes that of a disc in the plane perpendicular to $\OMEGA$. At this point, $\Q$ once again has two degenerate eigenvalues and so the distribution is uniaxial ($S = P$). However, the director is now perpendicular to $\OMEGA$, and $\Q$ describes disc like particles at the mesoscale. A subtle, but important, point is that the distribution spreads out in the plane perpendicular to $\OMEGA$. Thus the two eigenvectors corresponding to the two largest eigenvalues of $\Q$ are in this plane. At the core of the disclination, these two eigenvalues cross. 

In addition to $\T$ and $\OMEGA$, it is customary in two dimensions to define the ``orientation'' of a disclination line. The orientation is the phase origin of the the director as is encircles the line. In the normal plane, an angle $\phi$ may be defined that gives the azimuthal angle with respect to some reference axis. Further, the director at a reference angle $\phi_0$ is denoted $\n_0$ (see Eq. \eqref{eqn:nCore} for an approximate description of the director near the core). We will take $\phi_0 = 0$, and so the value of $\n_0$ will describe the orientation. For example, a $+1/2$ disclination is in the shape of a comet (see Fig. \ref{fig:3DDisclinations}). Taking $\phi$ to be the angle with respect to the $x$-axis, $\n_0 = \mathbf{\hat{x}}$ describes a $+1/2$ disclination with the head of the comet pointing in the $-\mathbf{\hat{x}}$ direction, while $\n_0 = \mathbf{\hat{y}}$ describes a $+1/2$ disclination pointing the opposite direction. Additionally it has been shown that the local orientation of disclination lines can be described by a series of tensors of ranks $1-3$ \cite{long21}. The rank $1$ tensor gives the polarity of $+1/2$ wedge points, the rank $2$ tensor gives the characteristic twist directions for a twist point, and the rank $3$ tensor gives the three primary directions associated with a trifold symmetric $-1/2$ wedge point (see Fig. \ref{fig:3DDisclinations}). For an arbitrary point, all three tensors may be used to fully describe the orientation. Although this is a more mathematical method for describing the orientation of disclination lines, for this work it will only be necessary to use $\n_0$ to describe disclination orientation.

\section{Disclination kinematics}\label{sec:LineKinematics}

In a two dimensional nematic, the order parameter $S \to 0$ as the core of a disclination is approached. This is similar to the case of superfluids and superconductors in which the order parameter goes to zero at vortex cores \cite{feynman55,abrikosov57,pismen99}. However, as shown in Sec. \ref{subsec:CoreStructure}, $S \neq 0$ at a disclination core in three dimensions. Instead, order goes from uniaxial to biaxial and back to uniaxial precisely at the core \cite{schopohl87,schimming21}. Both this lack of a singularity, and the geometric complexity of three dimensional nematic order near disclinations have prevented the extension of defect tracking methods to the case of three dimensional nematics. It is shown below that such tracking methods can be introduced in this case by focusing on the line $S=P$ in which there is a crossing of eigenvalues of the tensor $\mathbf{Q}$.

Consider a system with $N$ line disclinations, so that $\mathbf{R}_i(s)$ is the position of an element of line of the $i$th disclination for an arbitrary parametrization of the line. The macroscopic disclination density is \cite{liu92}
\begin{equation} \label{eqn:DiscDens}
    \bm{\rho}(\mathbf{r}) = \frac{1}{2}\sum_i \int \frac{d \mathbf{R}_i}{d s}\delta\left[\mathbf{r} - \mathbf{R}_i(s)\right]\,d s
\end{equation}
where the factor of $(1/2)$ arises from its topological charge, and the vector density $\bm{\rho}$ is directed along the line tangent $\T$. As discussed in Sec. \ref{subsec:CoreStructure}, the mesoscopic disclination core is diffuse (as also seen in experiments \cite{kim13,zhou17}), and the defect location $\mathbf{R}_{i}$ needs to be defined precisely. We define the location of the defect on the line $S = P$, for $S$ and $P$ defined by Eq. \eqref{eqn:QParam}.

At the core the order parameter only has three degrees of freedom: two that define the rotation vector, $\OMEGA$, and one that indicates the strength of ordering at the core, $S_C$. The director deformation satisfies $\OMEGA \cdot \n = 0$ on the plane normal to the disclination line \cite{friedel69} (close to the disclination core, the director remains in a single plane, the plane perpendicular to $\OMEGA$, as it encircles the core). Slightly away from the core on the normal plane to the line, but still within a diffuse core radius $a$, the order parameter is biaxial and has five degrees of freedom: three previously discussed describing the core, one for the orientation of the director (the dominant eigenvector in this biaxial region), and one for the difference between uniaxial and biaxial order, $\delta S = S - P$. In this region, the director may be written as
\begin{equation} \label{eqn:nCore}
    \n = \n_0 \cos\left(\frac{\phi - \phi_0}{2}\right) + \n_1 \sin\left(\frac{\phi - \phi_0}{2}\right)
\end{equation}
where $\{\n_0,\n_1,\OMEGA\}$ form an orthonormal triad and $\phi$ is the azimuthal angle in the normal plane with respect to a reference axis. Equation \eqref{eqn:nCore} is a useful approximation of the director near the core. Note that this relation is exact everywhere in the single elastic constant approximation, and for a single straight line defect with constant $\OMEGA$. In general, far field boundary conditions, the presence of other defects, or curvature of the defect line yield more complicated director configurations, especially as one moves further from the core. In the $\{\n_0,\n_1,\OMEGA\}$ basis, the tensor order parameter can be expressed as
\begin{multline} \label{eqn:SeparateQ}
    \Q(0 < |\mathbf{r} - \mathbf{R}|_{\perp} < a) = S_C \begin{pmatrix} \frac{2}{3} & 0 & 0\\ 0 & \frac{2}{3} & 0\\ 0 & 0 & -\frac{4}{3} \end{pmatrix} \\+ \delta S \begin{pmatrix} -\sin^2\frac{\phi}{2} & \frac{1}{2}\sin \phi & 0\\ \frac{1}{2} \sin\phi & - \cos^2 \frac{\phi}{2} & 0\\ 0 & 0 & 1\end{pmatrix}
\end{multline}
where $|\mathbf{r} - \mathbf{R}|_{\perp}$ indicates distance in the normal plane of the disclination, and $\delta S = S - P = 0$ at the core.
\begin{figure}
    \centering
    \includegraphics[width = \columnwidth]{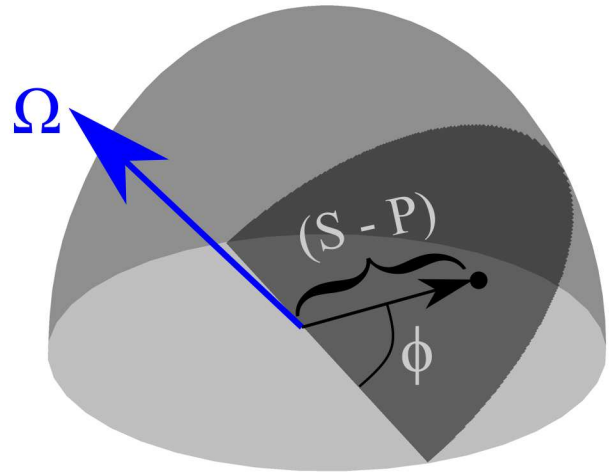}
    \caption{Schematic representation of the order parameter subspace near a disclination core. $\delta S = S - P$ acts as a radial coordinate while the director angle in the plane perpendicular to $\OMEGA$, $\phi$, acts as an azimuthal coordinate. The transformation from real space to this subspace may be viewed as a transformation to standard polar coordinates.}
    \label{fig:Subspace}
\end{figure}
Equation \eqref{eqn:SeparateQ} defines a two dimensional subspace schematically shown in Fig. \ref{fig:Subspace}. $\delta S$ is a radial coordinate on the subspace, while $\phi$ is the azymuth. We denote this space as $\Q_{\perp}$ as it is intuitively the part of order parameter space that is perpendicular to $\OMEGA$. 

We next introduce the Jacobian of the coordinate transformation from real space to this order parameter subspace. To accomplish this, we first note that if $\Q$ is parameterized as in Eq. \eqref{eqn:QParam}, with $\n$ given by Eq. \eqref{eqn:nCore}, $\bm{\hat{\ell}} = \OMEGA$, and $\mathbf{\hat{m}} = \OMEGA \times \n$, then the quantity
\begin{equation} \label{eqn:QGradQ}
    \hat{\Omega}_{\gamma}\varepsilon_{\gamma \mu \nu} Q_{\mu \alpha} \nabla Q_{\nu \alpha} = \frac{1}{2}(\delta S)^2\nabla \phi
\end{equation}
where summation of repeated indices is assumed. Given the identification of $\delta S$ and $\phi$ as polar coordinates on the subspace, (Fig. \ref{fig:Subspace}), and recalling that the Jacobian transformation from Cartesian to polar coordinates $(\rho, \theta)$ is $\nabla \times [(1/2)\rho^2 \nabla \theta]$, we may take the curl of Eq. \eqref{eqn:QGradQ} to define the transformation from the real space defect density to the density in order parameter space as,
\begin{align}\label{eqn:DensityQ}
    \bm{\rho}(\mathbf{r}) &= \delta[\Q_{\perp}] \OMEGA \cdot \mathbf{D}(\mathbf{r}) \\ \nonumber
    D_{\gamma i} &= \varepsilon_{\gamma \mu \nu} \varepsilon_{i k \ell} \partial_k Q_{\mu \alpha} \partial_{\ell} Q_{\nu \alpha}. 
\end{align}
This is the central result of this section that gives the transformation between the defect density in real and order parameter spaces (as indicated by the arguments of the Dirac delta functions in Eqs. \eqref{eqn:DiscDens} and \eqref{eqn:DensityQ}). Note that in taking the curl of Eq. \eqref{eqn:QGradQ} there should be three terms. However, the term $\sim \nabla \OMEGA \times (\Q \times \nabla \Q)$ goes to zero because the derivative of $\OMEGA$ is perpendicular to itself since $\OMEGA$ is a unit vector and $\Q \times \nabla \Q \propto \OMEGA$. The other term $\sim \OMEGA \cdot [\Q \times (\nabla \times \nabla \Q)]$ is likewise zero since $\Q$ is a non-singular quantity (i.e. the curl of the gradient is zero). We also note that $\OMEGA \cdot \mathbf{D} \propto \T$ (see Appendix \ref{append:DensityTensor}) as required since $\bm{\rho} \propto \T$. 

Equation (\ref{eqn:DensityQ}) is the analog of the transformation used to study point and line defects in $O(n)$ $n$-vector models \cite{halperin81,liu92,mazenko97,mazenko99}. Defects there are identified as zeros of the $n$-vector order parameter $\bm{\psi}$, and the map from real space to order parameter space leads to a defect density transformation of the form
\begin{equation*}
    \rho(\mathbf{r}) = \sum_i m_i \delta\left(\mathbf{r} - \mathbf{r}_i\right) = \delta \left[\bm{\psi}(\mathbf{r})\right] D(\mathbf{r})
\end{equation*}
where $\mathbf{r}_i$ is the location of the $i$th defect with charge $m_i$ and $D$ is the appropriate Jacobian. This formalism has been widely used to describe defects in superfluids, superconductors, XY models, and classical ferromagnets, to name a few. A similar approach has been used to study dislocations in solids using a phase field model of the lattice displacement \cite{audun18,skogvoll22}.

We conclude the discussion by noting that the tensor $\mathbf{D}$ that appears in Eq. \eqref{eqn:DensityQ} was introduced earlier in Ref. \cite{schimming22} in connection with the identification of tangent and rotation vectors of disclination lines. For completeness, we summarize those results in Appendix \ref{append:DensityTensor}, and discuss how they are related to the topology of disclination lines.

\subsection{Velocity of a disclination line}\label{subsec:Velocity}

The transformation \eqref{eqn:DensityQ} allows the derivation of the kinematic law of motion for the disclination line. The derivation is summarized here; details can be found in Appendix \ref{append:Velocity}. By taking the time derivative of the left hand side of Eq. \eqref{eqn:DensityQ}, given the definition of the density in \eqref{eqn:DiscDens}, one finds that 
$ \partial_t\rho_i = \partial_k\left(v_i\rho_k - v_k \rho_i\right) $ (Eq. \eqref{eqn:rhodtrho}), where  $\mathbf{v}(s) = d\mathbf{R}(s)/dt$ is the velocity of the line. On the other hand, the derivative of the right hand side can be computed by explicitly obtaining the time derivative of the disclination density tensor. The following conservative form is obtained
$$
\partial_t D_{\gamma i} = 2 \partial_{k} J_{\gamma i k}, \quad J_{\gamma i k} = \epsilon_{\gamma \mu \nu} \epsilon_{i k l} \partial_{t}Q_{\mu \alpha} \partial_{l} Q_{\nu \alpha},
$$
where $ J_{\gamma i k}$ is the disclination density current. This equation reflects the conservation of topological charge as the disclination density changes in order parameter space. Therefore one finds (also using the definition of the density \eqref{eqn:DensityQ}) that, \eqref{eqn:JEJ},
$$
    2 \hat{\Omega}_{\tau} J_{\tau i k}\delta\left[\Q_{\perp}\right] = \hat{\Omega}_{\gamma}\left(v_i D_{\gamma k} - v_k D_{\gamma i}\right)\delta\left[\Q_{\perp}\right],
$$
equality that applies only at the core of the disclination. This equation can be solved for the velocity by introducing an auxiliary tensor field $\mathbf{g}$, so that the velocity of a disclination line is,
\begin{align} \label{eqn:Velocity}
    \mathbf{v}(s) &= 2\left. \frac{\T\times(\OMEGA\cdot\mathbf{g})}{|\mathbf{D}|}\right|_{\mathbf{r} = \mathbf{R}(s)}\\
    g_{\gamma k} &= \varepsilon_{\gamma \mu \nu}\partial_t Q_{\mu \alpha}\partial_k Q_{\nu \alpha} \nonumber
\end{align}
where the tensor field $\mathbf{g}$ is related to the topological charge current (see Appendix \ref{append:Velocity}), and all quantities are computed at the disclination core. Note that the velocity is explicitly perpendicular to the tangent vector of the disclination, as expected. 

Equation \eqref{eqn:Velocity} is an exact kinematic relation between the velocity of a disclination line (defined as the line $S=P$) and the evolution equation of the tensor order parameter. Thus the equation is valid for any dynamic model of nematic evolution, be it simply diffusive relaxation, involve coupling to hydrodynamic transport, or be a model of an active phase. The details of the dynamic model are contained in the tensor $\mathbf{g}$, or more specifically, in its explicit dependence on $\partial_t \Q$. Another important property of Eq. \eqref{eqn:Velocity} is it only needs to be computed at the disclination core. This includes both tangent and rotation vectors, $\T$ and $\OMEGA$. This property will allow us to analytically predict defect motion in a variety of disclination configurations in subsequent sections by using an approximation for $\Q$ that is accurate close to the core.

Finally, we note that Eq. \eqref{eqn:Velocity} reduces to the expression derived in Ref. \cite{angheluta21} for the velocity of a disclination in a two dimensional nematic. In that case, by taking $\T = \mathbf{\hat{z}}$ and $\OMEGA = \pm \mathbf{\hat{z}}$, one finds
\begin{equation}
    v_i = \mp 4\frac{\varepsilon_{3 i k}\varepsilon_{3 \mu \nu} \partial_t Q_{\mu \alpha}\partial_k Q_{\nu \alpha}}{\varepsilon_{3\ell p}\varepsilon_{3\tau\xi}\partial_{\ell}Q_{\tau \beta}\partial_p Q_{\xi \beta}}.    
\end{equation}

\section{Linear core approximation. The Peach-Koehler force}\label{sec:Approximations}

Equation \eqref{eqn:Velocity} specifies the velocity of a disclination line in terms of derivatives of the order parameter only at the defect core. This is in general a complex problem that requires, in principle, the solution for the field $\Q$ everywhere. Considerable analytic progress can be made, however, by introducing the linear core approximation of Ref. \cite{long21}. This is similar to the linear core approximations made for vortices in superfluids and superconductors, or for the motion of dislocations in solids when modeled by a phase field \cite{audun18}. 

For the purposes of this section, when analyzing an arbitrary point on a disclination line, we will adjust our axes so the point of interest is located at $\mathbf{r} = 0$ and take the azimuthal angle $\phi = 0$ to coincide with the positive $x$ axis so that the tangent vector to the disclination line is $\T = \mathbf{\hat{z}}$. The linear core approximation is derived by first noting that in the uniaxial region away from the core $\Q$ may be written in terms of the vectors $\{\n_0,\n_1,\OMEGA\}$ in Eq. \eqref{eqn:nCore}:
\begin{multline}
    \Q = S_N\left[\frac{1}{6}\mathbf{I} - \frac{1}{2}\OMEGA \otimes \OMEGA + \frac{1}{2}\cos\phi\left(\n_0\otimes\n_0 - \n_1\otimes\n_1\right)\right.\\ \left. + \frac{1}{2}\sin\phi\left(\n_0\otimes\n_1 + \n_1\otimes\n_0\right)\right].
\end{multline}
Inside a diffuse core of radius $a$, $\Q$ is linearly interpolated by replacing $\cos\phi \to x/a$ and $\sin\phi \to y/a$, so that $\Q$ near the core is approximately given by
\begin{multline} \label{eqn:QApprox}
    \Q = S_N\left[\frac{1}{6}\mathbf{I} - \frac{1}{2}\OMEGA \otimes \OMEGA + \frac{x}{2a}\left(\n_0\otimes\n_0 - \n_1\otimes\n_1\right) \right. \\ \left.+ \frac{y}{2a}\left(\n_0\otimes\n_1 + \n_1\otimes\n_0\right)\right].
\end{multline}
As shown in Ref. \cite{long21}, this approximation for $\Q$ is quite good near the point where the eigenvalues cross and, remarkably, it is also fully biaxial in the region $0<\rho<a$, even though far from the core $\Q$ is purely uniaxial. We will use this approximation to make a number of predictions for the disclination velocity in several prototypical configurations by using it in conjunction with Eq. \eqref{eqn:Velocity}.

We focus first on simple diffusive relaxation of the tensor order parameter,
\begin{equation}
    \partial_t \Q = - \Gamma \frac{\delta F}{\delta \Q},
\end{equation}
where $F$ is the free energy. If $F$ has a functional derivative with non gradient terms that are analytic in $\Q$ at the disclination core (such as the Landau-de Gennes free energy or the model used here for numerics---see Sec. \ref{subsec:Implementation} and Appendix \ref{append:CompDetails}), then the velocity of the line does not depend on those terms. This follows from $g_{\gamma k} = \varepsilon_{\gamma \mu \nu} (\Q^n)_{\mu\alpha}\partial_k Q_{\nu \alpha} = 0$ when computed at the core for any power $n$. Thus one needs only focus on gradient terms from the elastic free energy. In the one elastic constant approximation we may write $\partial_t\Q \propto \nabla^2 \Q$ in Eq. \eqref{eqn:Velocity}. We will assume this gives the evolution of $\Q$ for the rest of the paper unless otherwise specified.

First, for a single, straight line disclination, Eq. \eqref{eqn:Velocity} predicts $\mathbf{v} = 0$, since $\nabla^2 \Q = 0$ at the core of the disclination. This is the correct stationary state for a single straight line disclination. However, as we will show, curvature in the disclination line, even if isolated, may result in a nonzero velocity.

One way to induce disclination motion is through an externally imposed distortion of the director field. The simplest case (and most relevant to interacting disclinations) is a small, nonuniform rotation of angle $\tphi(\mathbf{r})$ of the director field near the disclination core about an axis $\q$. In this case
\begin{equation}
    \n \to  \tn = \cos\tphi \n + \sin\tphi \left(\q \times \n\right) + \left(1 - \cos\tphi\right)\left(\q\cdot\n\right).
\end{equation}
We further assume that $\tphi$ is small near the core so that $\tn \approx \n + \tphi(\q \times \n)$. We then use Eq. \eqref{eqn:nCore} to express $\tn$ near the disclination core:
\begin{equation} \label{eqn:nTilde}
\tn = \cos\frac{1}{2}\phi \tn_0 + \sin\frac{1}{2}\phi \tn_1    
\end{equation}
where $\tn_0$ and $\tn_1$ are defined analogously to $\tn$.

We now have all of the pieces of our approximations to use Eq. \eqref{eqn:Velocity} to predict disclination motion. Combining the linear core approximation, Eq. \eqref{eqn:QApprox}, and the perturbed director near the core, Eq. \eqref{eqn:nTilde}, gives the final approximation for the perturbed $\Q$ near the core
\begin{multline} \label{eqn:FullQApprox}
    \Q \approx S_N\left[\frac{1}{6} \mathbf{I} - \frac{1}{2}\OMEGA\otimes\OMEGA + \frac{x}{2a}\left(\tn_0 \otimes \tn_0 - \tn_1 \otimes \tn_1\right) \right. \\ \left. + \frac{y}{2a}\left(\tn_0 \otimes \tn_1 + \tn_1 \otimes \tn_0\right)\right] 
\end{multline}
where $\tn_i \equiv \n_i + \tphi (\q \times \n_i)$.

Substituting Eq. \eqref{eqn:FullQApprox} into Eq. \eqref{eqn:Velocity} (details of this calculation are given in Appendix \ref{append:Analytic}) yields a simple expression for the velocity of the disclination line:
\begin{equation} \label{eqn:RotateMotion}
    \mathbf{v} = \left. -4\left(\q \cdot \OMEGA\right)\left(\T \times \nabla \tphi\right)\right|_{\rho=0},
\end{equation}
where we have expressed quantities in dimensionless units defined in Sec. \ref{subsec:Implementation}. In addition to giving the line velocity, it shows that, in particular, if the director is subjected to a small, \textit{nonuniform} rotation, a point on a disclination line will move if $\OMEGA\cdot\q \neq 0$ and if $\T$ at that point is not parallel to $\nabla \tphi$. We note that Eq. \eqref{eqn:RotateMotion} is the analog of the Peach-Koehler relation of dislocation mechanics \cite{peach50,pismen99}. The Peach-Koehler force applied to nematic liquid crystals was first introduced by Kl\'{e}man \cite{kleman83} and has been recently used to study disclination line motion in various scenarios \cite{long21,long22}. Our analysis here using Eq. \eqref{eqn:Velocity} represents an alternative derivation of this result that does not directly compare disclinations to dislocations in solids. Further, as we show in various examples in the next section, the above method may be generalized to obtain similar velocity equations for systems with twisted defect orientations and anisotropic elasticity.

\begin{figure*}
    \centering
    \includegraphics[width = \textwidth]{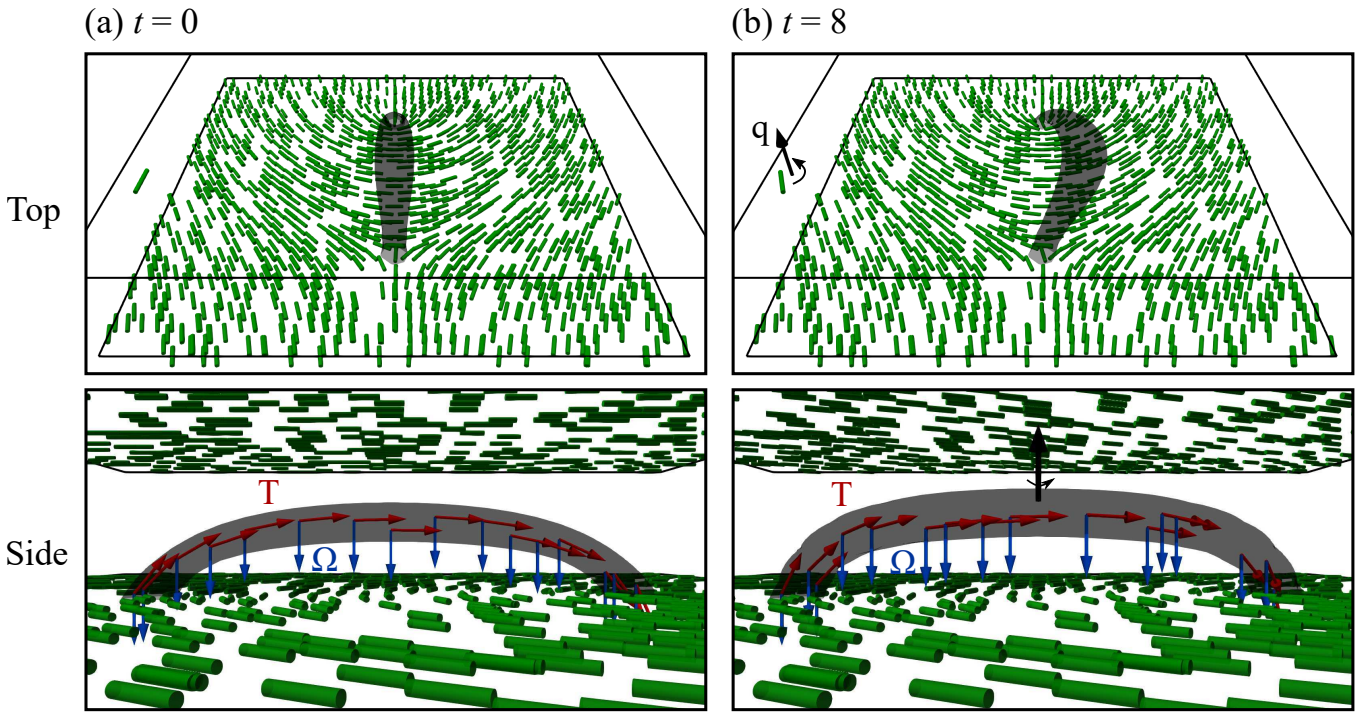}
    \caption{Disclination motion induced by externally imposed director rotation. The top row shows a top down view, while the bottom row shows a side view. (a) Disclination line at time $t=0$ formed between patterned $\pm 1/2$ wedge disclinations on the bottom boundary. The top boundary has director $\n = \mathbf{\hat{x}}$ fixed. (b) Same disclination line at $t = 8$ (computation units). For $t > 0$, the top boundary is changed to have director $\n = \cos\tphi_0 \mathbf{\hat{x}} + \sin\tphi_0 \mathbf{\hat{y}}$ with $\phi_0 = \pi/8$, simulating a rotation of the top boundary. This imposed rotation in turn imposes a stress on the configuration which results in motion of the disclination, predicted by Eq. \eqref{eqn:RotateMotion}. In the figures, the black arrows represent the axis of rotation $\q$, the red arrows indicate the tangent vector to the disclination $\T$, and the blue arrows indicate the rotation vector $\OMEGA$.}
    \label{fig:DiscPlates}
\end{figure*}

We have compared the result of \eqref{eqn:RotateMotion} with a numerical study of the motion of a disclination line between two plates with fixed nematogen orientation. The numerical details of the algorithm are given in Sec. \ref{subsec:Implementation} and in Appendix \ref{append:CompDetails}. The top plate boundary condition is $\n = \mathbf{\hat{x}}$, parallel to the plate, while the bottom plate boundary condition has $\n$ parallel to the plate but defining a $\pm 1/2$ disclination pair. Neumann boundary conditions for the director are specified on the lateral boundaries. The system is then allowed to relax to a stable configuration in which a three-dimensional disclination line forms connecting the $+1/2$ and $-1/2$ disclinations on the bottom plate. This state is shown in Fig. \ref{fig:DiscPlates}(a) along with its tangent and rotation vectors at various points, which are computed from $\mathbf{D}$ (see Appendix \ref{append:DensityTensor}). After the stable state is reached, the boundary condition on the top plate is instantaneously rotated so that the director is now given by $\cos\tphi_0 \mathbf{\hat{x}} + \sin\tphi_0 \mathbf{\hat{y}}$ where $\tphi_0 = \pi/8$. This constitutes a rotation about axis $\q = \mathbf{\hat{z}}$ and induces a gradient $\nabla \tphi \propto \mathbf{\hat{z}}$. As seen in Fig. \ref{fig:DiscPlates}(b), this induces motion in the disclination line, primarily at the midpoint where $|\T \times \nabla \tphi|$ is largest, in agreement with  Eq. \eqref{eqn:RotateMotion}.

\section{Analytical and numerical results}\label{sec:Results}

We address next how this method of analytically computing disclination velocities can be used to accurately predict the motion of multiple disclinations, as interacting disclinations behave as if they induce rotations in the local director field of one another.

\subsection{Numerical algorithm}\label{subsec:Implementation}
Here we briefly summarize the numerical method used in the previous and following sections. We model passive relaxation dynamics unless otherwise specified. That is, $\partial_t \Q = -\Gamma \delta F / \delta \Q$ where $\Gamma$ is a rotational diffusion coefficient. The free energy we choose may be written in two parts $F = \int \left[f_B(\Q) + f_e(\Q,\nabla\Q)\right]\, d\mathbf{r}$ where $f_B$ is the bulk part of the free energy density that describes the phase behavior of the nematic, while $f_e$ is an elastic free energy density that penalizes spatial variations. The bulk free energy we use is based on a singular Maier-Saupe potential originally analyzed by Ball and Majumdar and recently developed computationally \cite{ball10,schimming20,schimming21},
\begin{align} \label{eqn:BallMajumdar}
    f_B(\Q) &= -\kappa \Tr [\Q^2] - T \Delta s \\
    \Delta s &= -n k_B \int_{{\cal S}^2} p(\bm{\hat{\xi}}) \ln 4\pi p(\bm{\hat{\xi}})\, d\Sigma(\bm{\hat{\xi}}) \nonumber
\end{align}
where $\kappa$ is a phenomenological interaction coefficient, $\Delta s$ is the entropy density relative to the isotropic phase, $n$ is the number density of nematogens, $k_B$ is the Boltzmann constant, and $p(\bm{\hat{\xi}})$ is the constant temperature, orientational probability distribution. Note that the integral in $\Delta s$ is over the unit sphere. 

For the elastic free energy density we use
\begin{multline} \label{eqn:ElasticE}
    f_e(\Q,\nabla \Q) = L_1 \partial_k Q_{ij}\partial_k Q_{ij} + L_2 \partial_{j} Q_{ij} \partial_{k} Q_{ik} \\+ L_3 Q_{k\ell}\partial_k Q_{ij}\partial_{\ell} Q_{ij}
\end{multline}
where $L_i$ are elastic coefficients. For a uniaxial nematic, comparison of Eq. \eqref{eqn:ElasticE} and the Frank-Oseen elastic free energy yields the following mapping to the splay ($K_{11}$), twist ($K_{22}$), and bend ($K_{33}$) coefficients \cite{frank58,longa87,selinger19}:
\begin{align} \label{eqn:KConstants}
    K_{11} &= 2S_N^2 L_1 + S_N^2 L_2 - \frac{2}{3} S_N^3 L_3 \\ \nonumber
    K_{22} &= 2S_N^2 L_1 - \frac{2}{3} S_N^3 L_3  \\ 
    K_{33} &= 2S_N^2 L_1 + S_N^2 L_2 + \frac{4}{3} S_N^3 L_3. \nonumber
\end{align}
Note that $L_3 \neq 0$ is required to break the degeneracy $K_{11} = K_{33}$.

Equation \eqref{eqn:BallMajumdar} with a specific form of $p(\bm{\hat{\xi}})$ constrains $\Q$ to be given by Eq. \eqref{eqn:QDef} \cite{ball10,schimming20,schimming21}. This specific choice of bulk free energy makes the case of an anisotropic nematic numerically tractable \cite{schimming21}. In some of the presented cases we will use the one constant approximation, $L_2 = L_3 = 0$, and so the use of this free energy is not strictly necessary. In these cases, our qualitative results should be reproducible with a more common Landau-de Gennes free energy functional instead. Appendix \ref{append:CompDetails} provides details about the computational implementation of Eq. \eqref{eqn:BallMajumdar}.

In all cases we solve the equations of motion for $\Q$ by discretizing in space on a square (2D) or cube (3D) that is meshed with triangles or tetrahedra. We discretize in time by using a semi-implicit convex splitting algorithm \cite{wise09,zhao16,xu19}. The discretized matrix equations are then solved using the Matlab/C++ package FELICITY \cite{walker18} and the multigrid matrix equation solver AGMG \cite{notay10,napov11,napov12,notay12}. For all problems we use Neumann boundary conditions unless otherwise specified. 

Finally, all lengths are made dimensionless by the nematic correlation length $\xi = \sqrt{L_1/(n k_B T)}$, energies by $\xi^3 n k_B T$, and times by the nematic relaxation time scale $\tau = 1 / (\Gamma \xi^3 n k_B T)$. For all computations we set $L_1 = 0.5$ and $\Gamma = 1$ which set the length and time scale. This leaves the following dimensionless parameters for the system
\begin{equation}
    \frac{\kappa}{n k_B T},\quad \tilde{L}_2 = \frac{L_2}{L_1}, \quad  \tilde{L}_3 = \frac{L_3}{L_1}.
\end{equation}
We will always set $\kappa/(n k_B T) = 4$, which corresponds to a system in the nematic phase with $S_N = 0.6751$ \cite{schimming20} and we subsequently drop the tilde on $L_2$ and $L_3$. 

\subsection{Twisted defects in two dimensions}\label{subsec:TwistedDefects}
We first present results that apply to disclination pairs in two-dimensional systems. These systems have been thoroughly studied \cite{deGennes75,svensek02,toth02,vromans16,tang17,pearce21}, and as such, the results presented here are not new, but rather a reinterpretation and alternative derivation of previous results. We will also use the simpler two dimensional case to demonstrate how we apply Eq. \eqref{eqn:RotateMotion} to systems of interacting disclinations. 

In two dimensions, the director can be defined by its angle relative to the $x$-axis, $\phi$. In a system with $N$ disclinations, $\phi$ is given by
\begin{equation} \label{eqn:Nphi}
    \phi(x,y) = \sum_{i=1}^N m_i \arctan\left(\frac{y - y_i}{x - x_i}\right) + \phi_0
\end{equation}
$m_i = \pm 1/2$ is the charge of the $i$th disclination, $(x_i,y_i)$ is the position of the $i$th disclination, and $\phi_0$ is an overall phase factor determined by the orientations of all the defects. Eq. \eqref{eqn:Nphi} minimizes the one elastic constant Frank-Oseen free energy for a system constrained to have $N$ disclinations at points $(x_i,y_i)$. 

Note that, as disclinations are added to the system, the effect is to add a nonuniform rotation everywhere outward from the center of the disclination. From the perspective of the $j$th disclination, the rest of the disclinations add small, rotatory perturbations. Hence, we identify the field $\tphi$ in Eq. \eqref{eqn:RotateMotion} with $\phi(x,y) - \phi_j(x,y)$ where $\phi_j(x,y)$ is the angle of the director that is attributed only to the $j$th disclination. Then, using the two-dimensional version of Eq. \eqref{eqn:RotateMotion}, the velocity of the $j$th disclination in a two dimensional system of $N$ disclinations is
\begin{equation} \label{eqn:2DInteract}
    \mathbf{v}_j = 8 \sum_{i\neq j}m_i\frac{\mathbf{R}_j - \mathbf{R}_i}{|\mathbf{R}_j - \mathbf{R}_i|^2} 
\end{equation}
where $\mathbf{R}_i = (x_i,y_i)$. Eq. \eqref{eqn:2DInteract} is the well-known ``Coulomb-like'' interaction between disclinations in a nematic \cite{deGennes75}. Eq. \eqref{eqn:2DInteract} is traditionally derived by using the Frank-Oseen free energy, written in terms of disclination positions. The disclination kinematic law is an alternative method of deriving the same result. 

\begin{figure}
    \centering
    \includegraphics[width = \columnwidth]{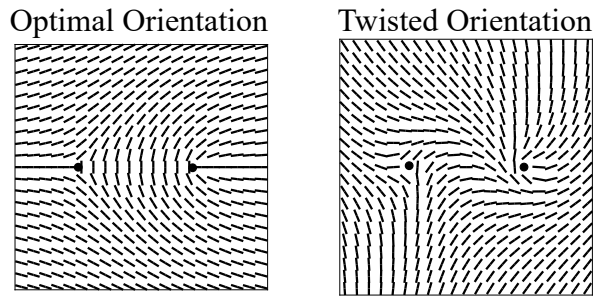}
    \caption{(Left) Director configuration for two oppositely charged two dimensional disclinations with ``optimal'' orientation. (Right) Director configuration for two oppositely charged two dimensional disclinations with ``twisted'' orientation. The angle between orientations $\delta\phi = \pi$.}
    \label{fig:OptimalAndTwist}
\end{figure}

Now consider the case of two oppositely charged disclinations. Equation \eqref{eqn:2DInteract} shows that the disclinations will annihilate moving along the line connecting the two disclination cores. Recently, the Frank-Oseen free energy has been minimized for the case of two disclinations fixed in space, but having arbitrary relative orientation $\delta \phi$ \cite{tang17}. The left panel of Fig. \ref{fig:OptimalAndTwist} shows the director field for the standard case in which the orientation between disclinations is ``optimal'' ($\delta \phi = 0$) while the right panel shows the case of ``twisted'' disclinations in which $\delta \phi = \pi$. One may think of this configuration as being formed by ``twisting'' one of the disclinations by an angle $2 \delta \phi$ relative to the other. Tang and Selinger, \cite{tang17}, showed that in the twisted case, and for defects sufficiently separated, the director angle is given by
\begin{multline} \label{eqn:TwistedDirector}
    \phi(x,y) = \frac{1}{2}\arctan\left(\frac{y - y_1}{x - x_1}\right) - \frac{1}{2}\arctan\left(\frac{y-y_2}{x-x_2}\right) \\ + \frac{\delta \phi}{2}\left[1 + \frac{\ln\left(|\mathbf{r} - \mathbf{R}_1|^2\right) - \ln\left(|\mathbf{r} - \mathbf{R}_2|^2\right)}{\ln\left(|\mathbf{r}_1 - \mathbf{r}_2|^2\right) - \ln\left(a^2\right)}\right] + \phi_0 
\end{multline}
where $a$ is the disclination core radius. If $\delta \phi = 0$ this case reduces to the optimal orientation case. The Frank-Oseen interaction energy of this configuration in terms of the distance between disclinations, $R = |\mathbf{R}_1 - \mathbf{R}_2| $ and the \lq\lq twistedness"  $\delta \phi$ of the configuration is \cite{tang17},
\begin{equation} \label{eqn:TwistedEnergy}
    \Delta F_{\text{FO}} = \frac{\pi K}{2} \ln\left(\frac{R}{a}\right) + \frac{\pi K \delta\phi^2}{2}\frac{\ln\left[R/(2a)\right]}{\ln\left[R/a\right]^2}
\end{equation}
where $K$ is the Frank-Oseen elastic constant in the one-constant approximation. Importantly, the energy only depends on the distance between the disclinations and the twistedness of the configuration. Thus, the force that one disclination exerts on the other is directed along the line segment that joins them.

\begin{figure}
    \centering
    \includegraphics[width = \columnwidth]{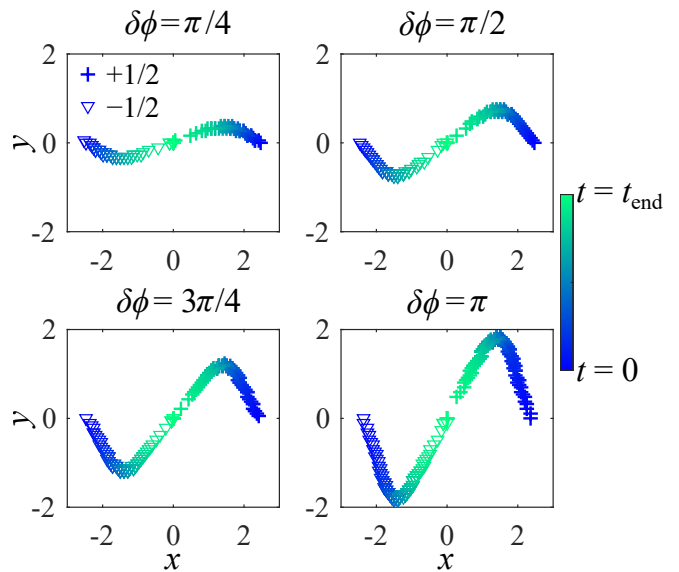}
    \caption{Trajectories of twisted disclinations for difference in orientations $\delta \phi = \pi/4,\,\pi/2,\,3\pi/4,\,\pi$. As $\delta \phi$ increases, the trajectories become more transverse. The transverse trajectories cannot be predicted from the energy of the configuration, yet the application of the kinematic velocity equation, Eq. \eqref{eqn:TwistedVelocity}, qualitatively captures the motion. In the figures, $+$ symbols represent the positions of the $+1/2$ disclination, while triangles represent the positions of $-1/2$ disclinations. For each case, the coloring indicates the time and is scaled from $t=0$ to $t=t_{\text{end}}$, the time at which the disclinations annihilate, which increases as $\delta \phi$ increases.}
    \label{fig:twistAnni}
\end{figure}

The disclinations, however, follow a more complex trajectory. As studied in Refs. \cite{vromans16,tang17,pearce21}, disclinations in twisted configurations have velocities with components transverse to the line segment that joins them. We show in Fig. \ref{fig:twistAnni} a few example trajectories numerically obtained. As the initial $\delta \phi$ is increased, the trajectories become more transverse, and straight line annihilation occurs only after the twisted distortion between the disclinations vanishes. Intuitively, the motion is due to a restoring torque, which drives the disclinations to rotate back to the optimal orientation. One way to rotate a disclination is by uniformly rotating all of the nematogens in the system. However, a uniform rotation in a system with two disclinations would only rotate both disclinations in the same direction, and hence would not reduce the relative twistedness between disclinations. Thus the nematogens must rotate locally, near each defect inhomogeneously. The only way this can be accomplished while maintaining continuity in the director field (at all points except the disclinations) is by moving the disclinations transverse to one another. 

This disclination behavior cannot be understood by energy minimization alone. Instead, the kinematic law, Eq. \eqref{eqn:Velocity}, can be used with $\tphi = \phi - \phi_1^{\text{singular}}$ where $\phi_1^{\text{singular}}$ is the part of Eq. \eqref{eqn:TwistedDirector} that is singular at $\mathbf{R}_1$. The predicted velocity of disclination $1$ is
\begin{equation} \label{eqn:TwistedVelocity}
    \mathbf{v}_1 = -2\left[\frac{1}{R} \mathbf{\hat{R}}_{12} - \frac{\delta\phi}{R\ln\left(R/a\right)}\left(\mathbf{\hat{z}}\times\mathbf{\hat{R}}_{12}\right)\right]
\end{equation}
where $\mathbf{\hat{R}}_{12} = (\mathbf{R}_1 - \mathbf{R}_2)/R$. The second term in Eq. \eqref{eqn:TwistedVelocity} is a transverse contribution that is proportional to $\delta \phi$. If $\delta \phi = 0$, the standard motion for optimally orientated disclinations follows. The velocity of disclination $2$ may also be derived in a similar manner and is precisely opposite to that of Eq. \eqref{eqn:TwistedVelocity}.

That the kinematic law gives a qualitatively correct velocity for twisted disclinations, whereas an over damped velocity proportional to the driving force does not, implies a tensorial effective mobility linking velocity and driving force that is topological of origin. From the perspective of the nematogens, the motion is geometrically constrained: nematogens must rotate locally while the director field remains continuous, and the only way to accomplish this is for the disclinations to move in the transverse direction. From the alternative perspective of defects being the primary dynamical objects, one may interpret this motion as a topological constraint that must be obeyed while energy minimization drives the relaxation.

\subsection{Motion in two dimensional anisotropic media}

We now consider two disclinations optimally oriented (not twisted relative to one another), though we relax the one constant approximation so that $K_{11} \neq K_{33}$. The director field around a single disclination when $K_{11} \neq K_{33}$ is known \cite{hudson89,zhou17,schimming20b}. However, for configurations involving two or more disclinations, the single defect solutions may not be simply superimposed because the Euler-Lagrange equations determining free energy minima are no longer linear \cite{dzyaloshinsky70}. Therefore, unlike the case of isotropic elasticity, the free energy cannot be computed analytically. The effect of one disclination on the other will be described by an unknown, local, inhomogeneous rotation of the director. From Eq. \eqref{eqn:Velocity} for the case of anisotropic elasticity (we assume $L_3 \neq 0$) the contribution to the dynamics of $\Q$ from the elastic free energy, Eq. \eqref{eqn:ElasticE}, is
\begin{multline} \label{eqn:anisotropicForce}
    \partial_t Q_{\mu \nu} = \nabla^2 Q_{\mu \nu} \\
    + L_3\left(-\partial_{\mu} Q_{ij} \partial_{\nu} Q_{ij} + 2\partial_i Q_{\mu \nu} \partial_j Q_{ij} + 2 Q_{ij} \partial_{\mu}\partial_{\nu}Q_{ij}\right).
\end{multline}

We now compute the velocity of the $+1/2$ disclination, disclination $1$, by noting that the director is rotated by a field $\tphi_2$, that is, the rotation from equilibrium caused by disclination $2$, the $-1/2$ disclination. We assume the same linear core approximation presented in Sec. \ref{sec:Approximations}, though the form of $\tphi_{2}$ is not known. Then by using the methods of Sec. \ref{sec:Approximations}, only the second term in the parentheses of Eq. \eqref{eqn:anisotropicForce} gives a nonzero velocity for disclination $1$,
\begin{multline} \label{eqn:AnisoV1}
    \mathbf{v}_1 = 4\left(\mathbf{\hat{z}} \times \nabla \tphi_2\right) \\- \frac{2 S_N L_3}{a}\left[\mathbf{\hat{x}}\cos2\phi_0 + \mathbf{\hat{y}}\sin2\phi_0\right]
\end{multline}
where $\phi_0$ is the overall phase of the configuration, defined in Eq. \eqref{eqn:Nphi}. The term proportional to $L_3$ does not depend on $\nabla \tphi_2$ within the linear core approximation, but depends instead on the orientation of the disclination through $\phi_0$. Of additional interest is in the computation of $\mathbf{v}_2$:
\begin{equation} \label{eqn:AnisoV2}
    \mathbf{v}_2 = 4\left(\mathbf{\hat{z}}\times\nabla\tphi_1\right)
\end{equation}
where $\tphi_1$ is the perturbation of the director from disclination $1$. Note that there are no terms proportional to $L_3$.

Equations \eqref{eqn:AnisoV1} and \eqref{eqn:AnisoV2} predict asymmetric motion of $\pm 1/2$ disclinations when $K_{11} \neq K_{33}$. Such asymmetric motion has been observed in previous numerical work \cite{toth02}. Additionally, the velocity equation reveals multiple sources of asymmetry. There is an explicit asymmetry in which the $+1/2$ disclination shows a biased motion towards its bend region if $L_3 > 0 $ or towards its splay region if $L_3 < 0$. The $-1/2$ disclination shows no such bias, as expected, since it is not polar. However, there is also an implicit asymmetry, since in general $\nabla \tphi_1 \neq -\nabla \tphi_2$, unlike the case for pairs of disclinations in the one-constant approximation in which $\nabla \tphi_1 = -\nabla \tphi_2$.

\begin{figure}
    \centering
    \includegraphics[width = \columnwidth]{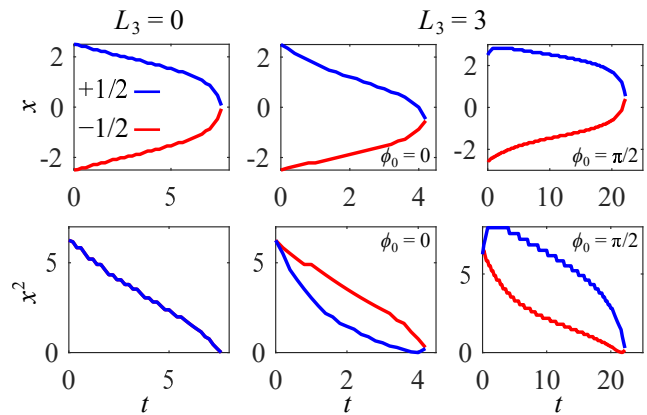}
    \caption{$\pm1/2$ disclination positions $x$ and $x^2$ versus time $t$ in computational units for $L_3 = 0$ and $L_3 = 3$. The case $L_3 = 3$ is further differentiated by initial conditions: $\phi_0 = 0$ and $\phi_0 = \pi/2$. When $L_3 > 0$ the bend elastic constant is larger than splay, leading to an asymmetry in $\pm 1/2$ disclination motion.}
    \label{fig:AnisoAnni}
\end{figure}

We have compared these results to a numerical solution of the time evolution of the $\Q$-tensor in which the system is initialized with two oppositely charged disclinations. For the computations we set $L_3 = 3$ and $\Delta t = 0.2$. We perform computations for two different cases, $\phi_0 = 0$ and $\phi_0 = \pi/2$. Fig. \ref{fig:AnisoAnni} shows plots of the position $x$ and $x^2$, as a function of time for the $+1/2$ and $-1/2$ disclinations for both anisotropic cases, as well as the case of an elastically isotropic system ($L_3 = 0$) for reference. We find that for $\phi_0=0$, the $+1/2$ disclination moves faster than the $-1/2$ disclination while the opposite occurs for the case $\phi_0 = \pi/2$. This is in agreement with Eq. \eqref{eqn:AnisoV1}. Additionally, the total time of annihilation is much smaller for the $\phi_0 = 0$ case than the $\phi_0 = \pi/2$ case. Further, the plots of $x^2$ are non-linear for the anisotropic cases, which is in contrast to case with isotropic elasticity when the velocity is given by Eq. \eqref{eqn:2DInteract}. The difference in disclination annihilation times can be understood by noticing that the $\phi_0 = 0$ configuration has bend deformation between the two disclinations and, hence, this configuration annihilates faster to remove the bend deformation, as the bend constant is larger than the splay constant when $L_3 > 0$. On the other hand, the $\phi_0 = \pi/2$ configuration contains splay between the disclinations resulting in slower motion. However, this intuitive argument does not account for asymmetry in disclination motion, which is accounted for qualitatively by Eqs. \eqref{eqn:AnisoV1} and \eqref{eqn:AnisoV2}.

\subsection{Disclination line interaction}\label{subsec:LineInteractions}
We turn our attention next to disclination lines in three dimensions with motion driven by their mutual interaction. It is not possible to evaluate the energy of configurations with many disclinations. Indeed, what constitutes an allowable configuration of many disclinations lines remains an open question. Nevertheless, we analyze two simple, yet interesting configurations: disclination line recombination, and disclination loop self-annihilation.

\subsubsection{Line recombination}

Disclination line recombination occurs when two disclinations meet at a point. They annihilate at this point and then continue to annihilate as separate lines. Here we calculate the velocity of disclination lines in a system of two straight lines with arbitrary tangent and rotation vectors. We set our coordinates so disclination $1$ is located at $\mathbf{R}_1(z) = (-R/2, 0, z)$ with tangent vector $\T_1 = \mathbf{\hat{z}}$. We further orient the coordinate system so that the closest point between the disclination lines lies on the $x$-axis and the tangent vector of disclination $2$ lies in the $yz$-plane so that $\mathbf{R}_2(z) = (R/2,|\mathbf{\hat{z}}\times \T_2|z,(\mathbf{\hat{z}}\cdot\T_2)z)$. The rotation vectors are assumed to be constant along the straight lines, but are otherwise arbitrary. Note that we use the coordinate $z$ as the parameter for both disclination lines.

\begin{figure}
    \centering
    \includegraphics[width = \columnwidth]{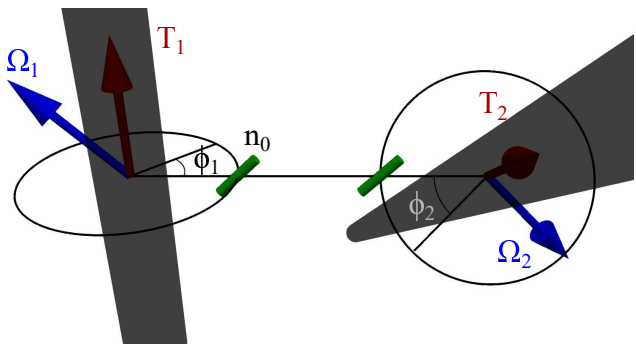}
    \caption{Schematic of the configuration for two straight line disclinations with constant rotation vectors $\OMEGA$. $\phi_1$ and $\phi_2$ are the azimuthal angles around each disclination defined such that $\phi_1 = \phi_2 = 0$ along the line segment that connects the closest points of the disclinations. The configuration is assumed to be such that $\n_0$ is the same for both disclinations.}
    \label{fig:TwoLineSketch}
\end{figure}

Unlike in two dimensions, an equation for the exact director field for multiple disclination lines is not known due to the nonlinearity of the Frank-Oseen free energy in three-dimensions \cite{frank58,selinger19}. However, we assume again that each disclination rotates the director field in its normal plane about its rotation vector, $\OMEGA$, and we can thus estimate the effect of one disclination on the other's local director field. To do this, we define two fields, $\phi_1(x,y,z)$ and $\phi_2(x,y,z)$ which give the azimuthal angle with respect to the normal planes of disclinations $1$ and $2$. We assume that $\phi_1$ and $\phi_2$ are both zero along the line segment that connects the nearest point of the disclination lines and that the disclination lines share $\n_0$ so that $\OMEGA_1 \cdot \n_0 = \OMEGA_2\cdot \n_0 = 0$. We show a schematic of this configuration in Fig. \ref{fig:TwoLineSketch}. 

To compute the velocity of disclination $1$ using Eq. \eqref{eqn:Velocity} we assume the director field near disclination $1$ is given by a small rotation of the director about the axis $\OMEGA_2$. This allows us to apply Eq. \eqref{eqn:RotateMotion} near the disclination with $\tphi = (1/2)\phi_2$ and $\q = \OMEGA_2$. This gives the velocity of disclination $1$ as a function of $z$,
\begin{multline} \label{eqn:DiscLineVelocity}
    \mathbf{v}_1(z) = -2\left(\OMEGA_1\cdot\OMEGA_2\right)\left[\frac{|\T_1\times\T_2|z}{R^2 + |\T_1\times \T_2|^2 z^2}\mathbf{\hat{y}} \right. \\ \left. + \frac{\left(\T_1\cdot\T_2\right)R}{R^2 + |\T_1\times \T_2|^2 z^2}\mathbf{\hat{x}}\right].
\end{multline}
Equation \eqref{eqn:DiscLineVelocity} gives several qualitative predictions about the motion of recombining disclination lines. First, the velocity is largest at $z = 0$, the closest point between the lines, and this point moves along the line segment connecting the closest points of the disclinations. Additionally, if the lines are not parallel, then there is a component of the velocity for points $z\neq 0$ that is transverse to the direction between disclinations. This component is odd in $z$, and thus indicates that non-parallel lines will rotate to become parallel. 

If we focus on the point $z = 0$ we find
\begin{equation} \label{eqn:ClosePointVel}
    \mathbf{v}_1(0) = 2\left(\OMEGA_1 \cdot \OMEGA_2\right)\left(\T_1\cdot\T_2\right)\frac{\mathbf{\hat{R}}_{12}}{R}
\end{equation}
so that the closest point does not move if the tangent vectors \textit{or} rotation vectors are perpendicular to each other. This motion was analyzed numerically in a previous work by us \cite{schimming22}, and it was found that this equation for the velocity of the closest points correctly predicts the scaling of numerical computations of annihilating disclinations. Further, we note that Eq. \eqref{eqn:ClosePointVel} is proportional to the force between two disclinations derived in Ref. \cite{tang17} by using an effective Peach-Koehler force between disclinations. Here, however, we do not integrate the force between two disclinations as is done in that work. Instead, Eq. \eqref{eqn:DiscLineVelocity} gives the velocity at all points along the disclination, predicting a non-uniform, rotating, motion.

We now expand upon our analysis by developing a simpler model for the time evolution of two important variables in the system: the distance between disclinations at their closest point, $R$, and the angle between tangent vectors at their closest point, $\psi$. That the disclinations rotate is important to their dynamics, since this rotation speeds them up as time goes on. Using Eq. \eqref{eqn:DiscLineVelocity} as well as the velocity for disclination $2$, which can be derived in a similar manner to the velocity of disclination $1$, we derive an equation for the time evolution of $R$ by noting that $dR/dt = (\mathbf{R}_1 - \mathbf{R_2})\cdot(\mathbf{v}_1 - \mathbf{v}_2)/R$. Additionally, the time evolution of $\psi$ may be derived as follows
\begin{align*}
    \frac{d}{dt}\left(\T_1\cdot\T_2\right) &= \frac{d}{dt}\cos \psi \\
   \Leftrightarrow \frac{d\T_1}{dt}\cdot\T_2 + \T_1\cdot\frac{d\T_2}{dt} &= -\sin\psi \frac{d\psi}{dt}
\end{align*}
with
\begin{equation*}
    \frac{d\T_i}{dt} = \frac{d}{dt}\frac{d\mathbf{R}_i}{dz} = \frac{d\mathbf{v}_i}{dz}.
\end{equation*}
Our simpler model for the dynamics of the closest points between disclinations is then given by two coupled, first order, differential equations
\begin{align}
    \frac{dR}{dt} &= \frac{4\left(\OMEGA_1\cdot\OMEGA_2\right)\cos\psi}{R} \label{eqn:dtR}\\
    \frac{d\psi}{dt} &= \frac{4\left(\OMEGA_1\cdot\OMEGA_2\right)\sin\psi}{R^2}. \label{eqn:dtPsi}
\end{align}
Equations \eqref{eqn:dtR} and \eqref{eqn:dtPsi} are, of course, an approximation that assumes (among others) that the disclinations remain straight, which is not the case in experiments and numerical calculations \cite{schimming22,zushi22}. Nevertheless, we note a few key predictions they make. First, as previously noted, if the rotation vectors are perpendicular, there should be no motion between the disclination lines, which has been predicted and found numerically previously \cite{long21,schimming22}. Further, if the disclinations are perpendicular and $\cos\psi = 0$ the distance between disclinations does not change. However, this does not mean that the disclinations do not move since, in this case, $\sin\psi = 1$ and so $\psi$ changes and the disclinations rotate. Additionally, the rotation rate is proportional to $1/R^2$ and so this rotation may take much longer if the disclinations are initially far apart.

\begin{figure}
    \centering
    \includegraphics[width = \columnwidth]{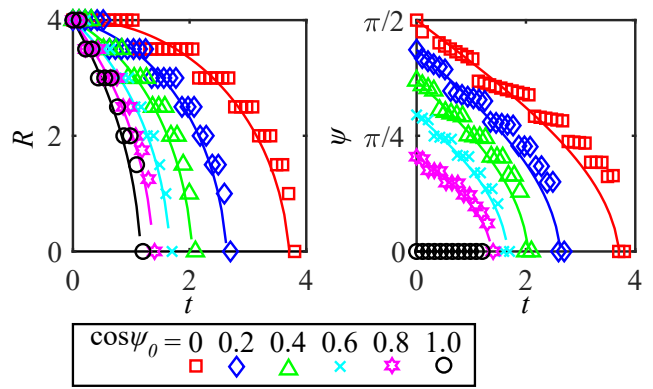}
    \caption{Disclination separation $R$ (left) and angle between tangent vectors $\psi$ (right) at the closest points between disclinations versus time during disclination recombination for several initial angles between disclinations. For all computations, the initial distance between disclinations is $R_0 = 4$. The points are data from $\Q$-tensor computations using the model of Sec. \ref{sec:Results} while the solid lines represent numerical solutions to Eqs. \eqref{eqn:dtR} and \eqref{eqn:dtPsi}. The time in the numerical solutions to Eqs. \eqref{eqn:dtR} and \eqref{eqn:dtPsi} is scaled so that the annihilation coincides with that of the $\Q$-tensor computations.}
    \label{fig:Recombination}
\end{figure}
 
We numerically solve Eqs. \eqref{eqn:dtR} and \eqref{eqn:dtPsi} by using a simple forward Euler method in which we take the time step $\Delta t = 0.1$, and we assume $\OMEGA_1\cdot\OMEGA_2 = -1$. We compare these solutions against full $\Q$-tensor computations of the model given in Sec. \ref{sec:Results}, setting $L_2 = L_3 = 0$ and $\Delta t = 0.1$. The three dimensional computations were performed on a standard tetrahedral mesh with $41\times41\times41$ vertices. In the computations we set $\OMEGA_1 = \mathbf{\hat{z}}$ and $\OMEGA_2 = -\mathbf{\hat{z}}$ and initialize the system so the initial distance between disclinations is $R_0 = 4$ with a range of initial tangent vectors so that $\cos\psi_0 \in [0,1]$. We track the tangent vectors of the disclinations using the $\mathbf{D}$ tensor (see Appendix \ref{append:DensityTensor}). In Fig. \ref{fig:Recombination} we plot $R$ and $\psi$ as a function of time for each initial condition used. In the plots, the solid lines are the Euler solutions to Eqs. \eqref{eqn:dtR} and \eqref{eqn:dtPsi} while the points are determined from the $\Q$-tensor computations. The jumps in data from the computations stem from the finite step-size of the mesh.

The solid lines in Fig. \ref{fig:Recombination} are not fits to the computational data; however the time is scaled so that the solutions of the model annihilate at the same time as the $\Q$-tensor computations. We find excellent agreement between the two methods, which highlights the power of the kinematic equation for disclinations in analyzing and predicting disclination motion since the differential equations \eqref{eqn:dtR} and \eqref{eqn:dtPsi} are much simpler, and faster to solve. We also note that the results of Ref. \cite{long21} predict that the force between two perpendicular disclinations should be zero, yet we find computationally, and predict analytically, that disclinations should still eventually annihilate due to a restoring torque between non-parallel disclinations.

\subsubsection{Loop self annihilation}

\begin{figure}
    \centering
    \includegraphics[width = \columnwidth]{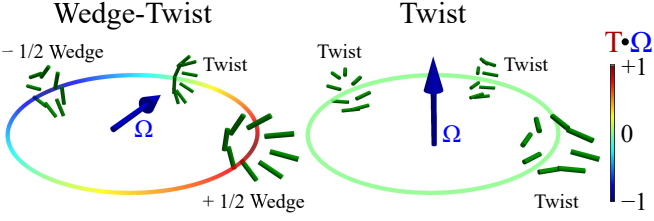}
    \caption{Sketches of wedge-twist and twist loop disclinations. The color indicates $\OMEGA \cdot \T$ along the loop while the cylinders depict the director at various points along the loop. These points are labeled by the type of elastic distortion present.}
    \label{fig:Loops}
\end{figure}

We now study the self annihilation of initially circular disclination loops in nematics. Axes are oriented so the center of the loop is at the origin and the loop lies in the $xy$-plane. We then work in standard cylindrical coordinates. Here we focus on disclination loops in which the rotation vector $\OMEGA$ is constant throughout. These are the so-called ``neutral'' disclination loops such that their total point defect charge is zero \cite{deGennes75,duclos20}. Thus, they may self annihilate to leave behind a uniform, defect-free, nematic texture. Additionally, there are two primary, geometric classifications of neutral disclination loops: ``twist'' disclinations in which $\OMEGA$ is perpendicular to the plane of the loop and hence $\OMEGA \cdot \T = 0$ everywhere; and ``wedge-twist'' disclinations in which $\OMEGA$ lies in the plane of the loop and $\OMEGA \cdot \T \in \left[-1,1\right]$ changes continuously along the loop. Sketches of both configurations are given in Fig. \ref{fig:Loops}.

To approximate $\Q$ near the core, we assume the director in each normal plane of the loop is given by
\begin{equation}
    \n = \cos\left(\frac{1}{2}\phi_1 + \frac{1}{2}\phi_2\right)\n_0 + \sin\left(\frac{1}{2}\phi_1 + \frac{1}{2}\phi_2\right)\n_1
\end{equation}
where
\begin{align*}
    \phi_1(\rho, \theta, z) &= \arctan\left(\frac{z}{R - \rho}\right) \\
    \phi_2(\rho, \theta, z) &= \arctan\left(\frac{z}{R + \rho}\right)
\end{align*}
where $R$ is the loop radius, and $\phi_2$ represents the rotation of the director field from the opposite side of the loop. The introduction of cylindrical coordinates slightly modifies the linear approximation of $\Q$ near the disclination core, which is now given by
\begin{multline}
    \Q \approx S_N\left[\frac{1}{6}\mathbf{I} - \frac{1}{2}\OMEGA\otimes\OMEGA + \frac{R - \rho}{2a}\left(\tn_0 \otimes \tn_0 - \tn_1 \otimes \tn_1\right) \right. \\
    \left. + \frac{z}{2a}\left(\tn_0\otimes\tn_1 + \tn_1 \otimes \tn_0\right)\right]
\end{multline}
where $\tn$ is defined as in Section \ref{sec:Approximations} and $\tphi = \phi_2$. In order to use the kinematic law, Eq. \eqref{eqn:Velocity}, to obtain the velocity of the disclination loop, we cannot just apply Eq. \eqref{eqn:RotateMotion} since the loop is not straight, and the curvature of the loop will add to the velocity (i.e. $\nabla^2 \rho \neq 0$, etc.). Using the same method as laid out in Sec. \ref{sec:Approximations} and taking the tangent vector to the loop to be $\T = \bm{\hat{\theta}}$ we find that the velocity is 
\begin{equation} \label{eqn:LoopVelocity}
    \mathbf{v} = -\frac{3}{2R}\bm{\hat{\rho}}.
\end{equation}

The velocity in Eq. \eqref{eqn:LoopVelocity} does not depend on the rotation vector $\OMEGA$. This is expected since, within the one-constant approximation, the Frank-Oseen free energy of a disclination loop does not depend on $\OMEGA$. Additionally, the velocity predicts that the loop shrinks at the same rate everywhere until it annihilates itself. The radius of the loop is predicted to scale as $R^2 \sim -t$ which was checked numerically in a previous work and has been observed in experiments \cite{schimming22,zushi22}. We reiterate that this method does not require an integration about the loop to predict the velocity. Rather we simply approximate the configuration $\Q$ at points along the loop. Finally, we note in a previous analysis (\cite{schimming22}) we did not include the contribution of $\tphi = \phi_2$ to the velocity (i.e. interaction with the opposite side of the loop) and only included the contribution due to disclination curvature. When including the contribution of $\phi_2$, the velocity is predicted to be larger by a factor of $3/2$. Further, comparing to Eqs. \eqref{eqn:dtR} and \eqref{eqn:dtPsi} for the case of parallel disclinations, a loop of radius equal to half the distance between disclinations is predicted to annihilate in $2/3$ the amount of time. Comparing $\Q$ tensor computations of disclination lines and loops shows that parallel disclinations initially separated by $R_0 = 5$ annihilate in $21$ time-steps ($\Delta t = 0.1$), while a loop disclination with initial diameter $2R_0 = 5$ annihilates in $12$ time-steps which is close to the factor of $2/3$ predicted by the equations.

We conclude this section with one more example of a self-annihilating disclination loop. We consider anisotropic elasticity and set $L_2 > 0$ to describe the case of larger (but still equal) bend and splay elastic constants relative to the twist constant [see Eq. \eqref{eqn:KConstants}]. With $L_2 \neq 0$ in Eq. \eqref{eqn:ElasticE}, the assumed time dependence of $\Q$ will change. We find now
\begin{equation}
    \partial_t Q_{ij} = \partial_k\partial_k Q_{ij} + L_2 \partial_i \partial_k Q_{jk}.
\end{equation}

In computing the velocity via the kinematic equation, we now choose a specific $\OMEGA = \mathbf{\hat{x}}$. Due to elastic anisotropy, the velocity will depend on the director configuration of the loop. If the disclination is a twist loop, all of the elastic deformation around the loop is twist. In this case, the disclination will still have velocity given by Eq. \eqref{eqn:LoopVelocity} since there is no change to the twist constant upon increasing $L_2$. Instead we focus on the case of a wedge-twist loop in which $\pm 1/2$ wedge and twist type deformations are represented along the loop (Fig. \ref{fig:Loops}).

For a circular wedge-twist loop disclination with $L_2 \neq 0$, we find that the velocity of the loop is
\begin{equation} \label{eqn:LowTwistLoop}
    \mathbf{v}(\theta) = -\frac{1}{2R}\left[3 + L_2\left(4 + 4\cos^2\theta - 2\sin\theta\right)\right]\bm{\hat{\rho}}
\end{equation}
where $\theta$ is the azimuthal angle with respect to the $x$-axis. For the loop under consideration, the twist portions of the loop occur at $\theta = 0,\,\pi$ while the $+1/2$ wedge portion occurs at $\theta = \pi/2$ and the $-1/2$ wedge portion occurs at $\theta = 3\pi/2$ (see Fig. \ref{fig:Loops}). For $L_2 > 0$, Eq. \eqref{eqn:LowTwistLoop} shows significant asymmetry in the evolution of the loop. The fastest sections of the loop turn out to be the twist sections, which is the opposite result to that obtained for straight, parallel line disclinations. For $L_2 > 0$, the bend and splay constants increase while the twist constant remains the same, and hence straight wedge ($\OMEGA\cdot\T = \pm 1$) disclinations will annihilate faster (because they cost more elastic energy) than straight twist ($\OMEGA\cdot\T = 0$) disclinations. Therefore it is the coupling to disclination curvature that causes this asymmetry. 

Additionally, there is asymmetry predicted in the wedge sections due to the $\sin\theta$ term in Eq. \eqref{eqn:LowTwistLoop}. This predicts that the $-1/2$ wedge section moves faster than the $+1/2$ wedge section if $L_2 > 0$. This is also against our intuition since the splay and bend constants are still equal for this set of parameters, and so for straight, parallel lines these two wedge defects would still annihilate symmetrically. We also note that Eq. \eqref{eqn:LowTwistLoop} holds regardless of the choice of $\n_0$ in the $yz$ plane.

\begin{figure}
    \centering
    \includegraphics[width = \columnwidth]{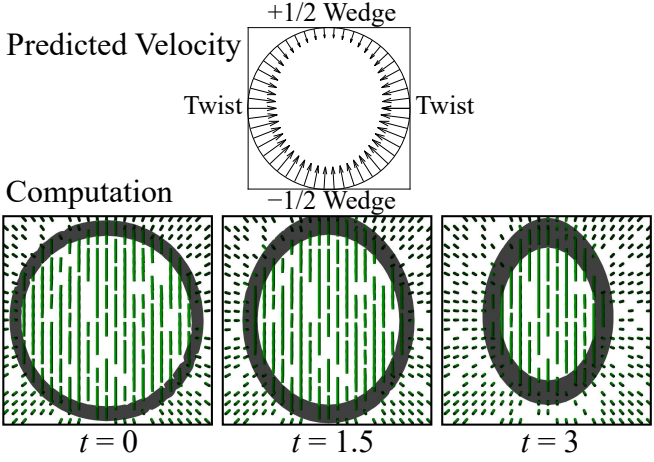}
    \caption{Annihilation of an initially circular wedge-twist loop disclination with $L_2 > 0$. The top panel shows the predicted velocity calculated in Eq. \eqref{eqn:LowTwistLoop}. The bottom panels show the configuration in the computation at three different times. The cylinders represent the nematic director, while the contours depict the extent of the disclination loop where $S = 0.3S_N$.}
    \label{fig:L2LoopAnni}
\end{figure}

We examine the predictions of Eq. \eqref{eqn:LowTwistLoop} through computation with the full $\Q$-tensor equations. For this computation, all parameters are the same as previous computations, except we now set $L_2 = 2$ which corresponds to a ratio of splay (or bend) to twist $K_{11}/K_{22} = 2$. Figure \ref{fig:L2LoopAnni} shows the predicted velocity alongside several time slices of the computational results. The predicted velocity very accurately captures the evolution of the loop, particularly at early times when the loop is circular. In the computation, the twist sections of the loop disclination move fastest and there is asymmetry between the $+1/2$ and $-1/2$ wedge sections. We note that recent experiments on systems of line and loop disclinations have not reported this asymmetry in loop annihilation, even though the experimental system has a twist constant approximately an order of magnitude smaller than the splay and bend constants \cite{zushi22}. This may indicate that all observed loop disclinations were twist type. This suggests that analytical calculations using the kinematic velocity law may be used to classify disclination types in experimental systems in which the material parameters are known.

\subsection{Defect sorting with external fields and flows}\label{subsec:ExternalFieldsFlows}
The final application of the kinematic law we examine is the motion of disclination lines under external fields or applied flows. These are two common situations studied in experiments on nematics, and are important in technological and biological applications as well as in the context of active nematics \cite{cladis87,vella05,biscari05,yeh09,lazo14,wang14,peng15,guillamat16,peng18,doo18,baza20,zhang21}.

\subsubsection{External Fields}

We first consider an external field that couples to the tensor order parameter free energy as  $f_H = -\chi\mathbf{H}^T \Q \mathbf{H}$ where $\mathbf{H}$ , the external field, can be an electric or magnetic feld \cite{deGennes75}, and $\chi$ is the susceptibility. If $\chi > 0$ this energy is minimized when $(\n \cdot \mathbf{H})^2 = |\mathbf{H}|^2$, that is, when the director aligns or anti-aligns with the field. For our purposes, the contribution to the time dependence of $\Q$ arising from this coupling is
\begin{equation}
    \partial_t Q_{ij} = \chi H_i H_j.
\end{equation}
The resulting velocity of a single, straight line disclination with constant $\OMEGA$,  assuming that the line lies along the $z$-axis and that it is oriented so $\phi_0=0$ corresponds to the positive $x$-axis, is
\begin{multline} \label{eqn:HVelocity}
    \mathbf{v} = \frac{\chi |\mathbf{H}|^2 a}{S_N}\left[\left((\mathbf{\hat{H}}\cdot\n_0)^2 - (\mathbf{\hat{H}}\cdot\n_1)^2\right)\mathbf{\hat{x}} \right. \\ \left.+ 2(\mathbf{\hat{H}}\cdot\n_0)(\mathbf{\hat{H}}\cdot\n_1)\mathbf{\hat{y}}\right]
\end{multline}
where $\mathbf{\hat{H}}$ is the unit vector in the direction of the applied field. If $\mathbf{H}$ is in the direction of $\OMEGA$, Eq. \eqref{eqn:HVelocity} shows there will be no motion of the disclination. Thus, we may limit further analysis to two dimensions and we will set $\OMEGA = \pm \mathbf{\hat{z}}$ and $\n_0 = \mathbf{\hat{x}}$. Note that if $\OMEGA \neq \mathbf{\hat{z}}$ then the following analysis holds for $\mathbf{H}$ in the plane perpendicular to $\OMEGA$. If we let $\mathbf{\hat{H}} = (\cos \beta,\sin\beta,0)$, Eq. \eqref{eqn:HVelocity} reduces to
\begin{equation} \label{eqn:HVel2D}
    \mathbf{v} = -\frac{\chi |\mathbf{H}|^2 a}{S_N}\left[\cos 2\beta \mathbf{\hat{x}} + 2 m \sin 2 \beta \mathbf{\hat{y}}\right]
\end{equation}
where $m = \pm 1/2$ is the charge of the effective two-dimensional disclination.

Equation \eqref{eqn:HVel2D} shows that if the field is aligned or anti-aligned with $\mathbf{\hat{x}}$ the disclination will move in the $-\mathbf{\hat{x}}$ direction. On the other hand, if the field is aligned or anti-aligned with $\mathbf{\hat{y}}$ the disclination will move in the $+\mathbf{\hat{x}}$ direction. This behavior is predicted to be independent of the charge of the disclination. However, if $\mathbf{\hat{H}}$ is skewed from these two alignments there is a predicted component of the velocity along the $\pm \mathbf{\hat{y}}$ direction, which depends on the charge $m$. Thus for disclinations that are oriented in the same direction, a skewed field will deflect oppositely charged disclinations in opposite directions.

\begin{figure}
    \centering
    \includegraphics[width = \columnwidth]{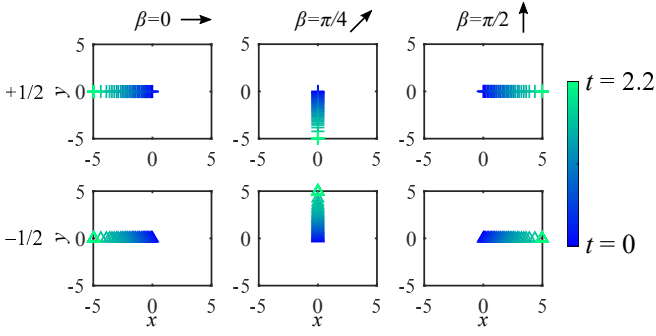}
    \caption{Disclination motion in the presence of an external field $\mathbf{H} = |\mathbf{H}|\left(\cos\beta,\sin\beta\right)$. The top panels show the trajectories of $+1/2$ wedge disclinations while the bottom panels show the trajectories of $-1/2$ wedge disclinations. The color indicates the time (in computational units).}
    \label{fig:HFieldDefects}
\end{figure}

We demonstrate this behavior by computing time dependent configurations in a two dimensional nematic with in plane field $\mathbf{H}$. For the computations we set $\Delta t = 0.5$, $\chi = 1$, and $|\mathbf{H}| = 0.5$. In Fig. \ref{fig:HFieldDefects} we show trajectories for $\pm 1/2$ disclinations with $\n_0 = \mathbf{\hat{x}}$ and $\beta = 0, \pi/4, \pi/2$. As shown in the figure, we find that Eq. \eqref{eqn:HVel2D} correctly predicts the direction of motion for the disclinations. In particular, when $\beta = \pi/4$ the motion of the $\pm 1/2$ disclinations is opposite one another, and hence the effect of this applied skewed field is to sort the disclinations by topological charge.

We may also compute the effect of external fields on three dimensional disclination loops. Applying the kinematic equation as we did in the previous section, assuming $L_2 = 0$, we find
\begin{multline} \label{eqn:HLoop}
    \mathbf{v} = \frac{\chi |\mathbf{H}|^2 a}{S_N}\left[\left((\mathbf{\hat{H}}\cdot\n_0)^2 - (\mathbf{\hat{H}}\cdot\n_1)^2\right)\bm{\hat{\rho}} \right. \\ \left.- 2(\mathbf{\hat{H}}\cdot\n_0)(\mathbf{\hat{H}}\cdot\n_1) \mathbf{\hat{z}}\right] - \frac{3}{2R}\bm{\hat{\rho}}
\end{multline}
where the second term is the self annihilation term derived above in Eq. \eqref{eqn:LoopVelocity}. Equation \eqref{eqn:HLoop} predicts a multitude of differing trajectories depending on $\OMEGA$ and $\n_0$. For the case of $\OMEGA = \mathbf{\hat{z}}$, a pure twist loop, only field components in the plane of the disclination loop are predicted to affect it. In this case, interestingly, if the field is directed along $\n_0$ Eq. \eqref{eqn:HLoop} predicts an unstable equilibrium radius of the loop. If the radius is smaller than the unstable length, the loop will self-annihilate. However, if the radius is larger than this length, the field will induce a continued growth in the loop. Additionally, if the field is skewed between the directions $\n_0$ and $\n_1$, Eq. \eqref{eqn:HLoop} predicts the disclination loop will deflect along the $\mathbf{\hat{z}}$ direction, depending on $\n_0$ and $\n_1$, that is, depending on the geometric properties of the loop.

\begin{figure}
    \centering
    \includegraphics[width = \columnwidth]{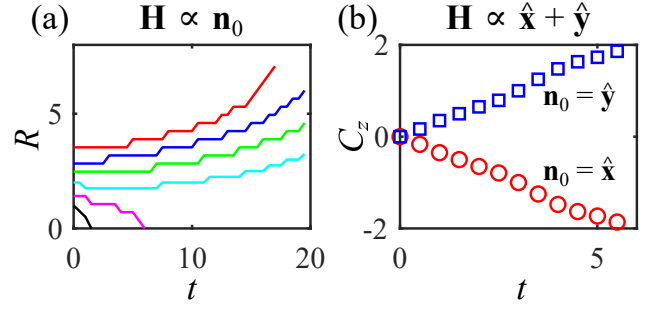}
    \caption{Computed motion of twist disclination loops ($\OMEGA = \mathbf{\hat{z}})$ in the presence of an external field $\mathbf{H}$. (a) Loop radius $R$ versus time $t$ for various initial radii when $\mathbf{H}$ is in the direction $\n_0$. Above an initial radius $R_0 \approx 2$ the loop grows indefinitely instead of self annihilating. (b) $z$-coordinate of the center of a twist disclination loop $C_z$ versus $t$ when $\mathbf{H} \propto \mathbf{\hat{x}} + \mathbf{\hat{y}}$. Two cases are shown: when $\n_0 = \mathbf{\hat{x}}$ the loop moves in the $-\mathbf{\hat{z}}$ direction while the opposite occurs when $\n_0 = \mathbf{\hat{y}}$.}
    \label{fig:HFieldLoops}
\end{figure}

We have also tested this prediction by solving the full $\Q$-tensor equations with $\mathbf{H} = 0.65 \n_0$ ($\n_0 = \mathbf{\hat{x}}$) and the rest of the parameters the same as above, only now in three dimensions for a pure twist loop. Figure \ref{fig:HFieldLoops}(a) shows the computed radius as a function of time for various initial radii. At $R_C \approx 2 $ we find the loop changes behavior from shrinking to growing indefinitely as predicted  by Eq. \eqref{eqn:HLoop}. From the computations, we estimate $a \approx 0.75$, which leads to a predicted critical radius from Eq. \eqref{eqn:HLoop} of $R_C \approx 2.1$, very close to our computational result.

Additionally, Fig. \ref{fig:HFieldLoops}(b) shows the $z$ coordinate of the center of a loop $C_z$ as a function of time for cases in which $\mathbf{H} \propto \mathbf{\hat{x}} + \mathbf{\hat{y}}$ with $\n_0 = \mathbf{\hat{x}}$ and $\n_0 = \mathbf{\hat{y}}$. As evidenced by the figure, the loop coherently moves down (up) along the $z$-axis when $\n_0 = \mathbf{\hat{x}}$ ($\n_0 = \mathbf{\hat{y}}$), as predicted by Eq. \eqref{eqn:HLoop}.

External fields have been shown to effectively identify disclinations either by topological or geometric content. Thus, these simple analytical results may lead to a number of applications in which different types of disclinations correspond to different active nematic or biological motifs. Colloidal particles, or differing cell types, have been shown to preferentially accumulate at regions of differing topological or geometric character, and hence external fields may allow particle and cell sorting \cite{gu00,ravnik07,alama16,genkin17,copenhagen21}. We note that the velocities given here [particularly the two dimensional Eq. \eqref{eqn:HVel2D}] could have also been predicted from the energy directly, since the effect of the field is to align the director. However, having the analytical tool is useful for more complex scenarios where energy methods may not be analytically viable.

\subsubsection{Shear Flow}

The motion of a disclination in an imposed flow $\mathbf{u}$ is studied next. We assume the time dependence of $\Q$ is given by the Beris-Edwards model \cite{beris94}:
\begin{multline} \label{eqn:BerisEdwards}
    \partial_t \Q = - (\mathbf{u}\cdot \nabla) \Q + \lambda\left[\mathbf{E}\Q + \Q \mathbf{E} + \frac{2}{3}\mathbf{E} \right. \\ \left. - 2\left(\Q + \frac{1}{3}\right)(\Q : \nabla\mathbf{u})\right] + \left[\mathbf{W},\Q\right] - \frac{\delta F}{\delta \Q}
\end{multline}
where $2 \mathbf{E} = \nabla \mathbf{u} + \nabla \mathbf{u}^T$ is the strain rate tensor, $2\mathbf{W} = \nabla \mathbf{u} - \nabla \mathbf{u}^T$ is the vorticity tensor, $\lambda$ is the ``tumbling'' parameter related to the tendency for the nematogens to align with shear \cite{leslie92,vanhorn00}, and $\left[\cdot,\cdot\right]$ is the commutator of two tensors. We choose this model because it is commonly employed in computational studies of active nematics in which disclinations play a primary role \cite{yeomans16}.

First, consider the contribution of the first term on the right hand side of Eq. \eqref{eqn:BerisEdwards}. This is the traditional advection term and is the only nonzero term in the case of uniform flow. The kinematic velocity equation for disclinations gives, for any segment of disclination with tangent vector $\T$,
\begin{equation} \label{eqn:advect}
    \mathbf{v} = 2\T \times \left(\mathbf{u} \times \T\right) + ~ {\rm relaxation},
\end{equation}
where \lq\lq + relaxation" stands for terms already discussed attributed to the diffusive relaxation of the disclination (for the rest of this section we will omit these terms though it is understood they still contribute to disclination motion). The right hand side of Eq. \eqref{eqn:advect} may also be written as $2[\mathbf{u} - (\mathbf{u}\cdot\T)\T]$, indicating that the flow in this case simply advects the disclination in the direction perpendicular to $\T$, with the speed reduced by the amount that $\T$ and $\mathbf{u}$ overlap. This is the expected result, and has been shown for two-dimensional disclinations already \cite{angheluta21}.

To study the effect of a more complicated---yet highly relevant---imposed flow, we turn our attention to an imposed shear flow,
\begin{equation}
    \mathbf{u}(\mathbf{r}) = \gamma \left(\mathbf{r}\cdot \mathbf{\hat{w}}\right)\mathbf{\hat{u}}
\end{equation}
where $\gamma$ is the shear rate, $\mathbf{\hat{u}}$ is the direction of the flow, and $\mathbf{\hat{w}}\cdot\mathbf{\hat{u}} = 0$. In this case, the tensors $\mathbf{E}$ and $\mathbf{W}$ are nonzero. However, because we only need to compute $\partial_t \Q$ at the disclination core to apply Eq. \eqref{eqn:Velocity}, it is easy to show that all terms in Eq. \eqref{eqn:BerisEdwards} that multiply $\Q$ will give zero in the computation of the disclination velocity. This is analogous to the case presented in Sec. \ref{sec:Approximations} in which terms involving $\Q^n$ for some power $n$ do not contribute to the disclination velocity. Thus, the only term we must consider is the flow aligning term, $(2\lambda/3)\mathbf{E}$. For the shear flow given,
\begin{equation}
    \mathbf{E} = \frac{\gamma}{2}\left(\mathbf{\hat{w}}\otimes\mathbf{\hat{u}} + \mathbf{\hat{u}}\otimes\mathbf{\hat{w}}\right).
\end{equation}

For a straight line disclination with constant $\OMEGA$ and $\T = \mathbf{\hat{z}}$, the flow contribution to the line velocity is
\begin{multline} \label{eqn:ShearLine}
    \mathbf{v} = 2\mathbf{\hat{z}}\times\left(\mathbf{u}\times\mathbf{\hat{z}}\right) \\+ \frac{2\lambda \gamma a}{3 S_N}\biggl[\Bigl(\left(\w\cdot\n_1\right)\left(\mathbf{\hat{u}}\cdot\n_1\right) - \left(\w\cdot\n_0\right)\left(\mathbf{\hat{u}}\cdot\n_0\right)\Bigr)\mathbf{\hat{x}}  \\  - \Bigl(\left(\w\cdot\n_0\right)\left(\mathbf{\hat{u}}\cdot\n_1\right) + \left(\w\cdot\n_1\right)\left(\mathbf{\hat{u}}\cdot\n_0\right)\Bigr)\mathbf{\hat{y}}\biggr].
\end{multline}
Equation \eqref{eqn:ShearLine} is similar to the velocity resulting from an imposed field, Eq. \eqref{eqn:HVelocity}, except there are two important directions instead of one. In particular, Eq. \eqref{eqn:ShearLine} predicts that if either the flow direction, $\mathbf{\hat{u}}$, or the shear direction, $\mathbf{\hat{w}}$, are parallel to $\OMEGA$, the only contribution to the disclination velocity from the flow will be advection. 

An illustrative example is when $\OMEGA = \pm \mathbf{\hat{z}}$, $\n_0 = \mathbf{\hat{x}}$, $\mathbf{\hat{u}} = \mathbf{\hat{x}}$, and $\w = \mathbf{\hat{y}}$. This is the effective two-dimensional case for $\pm 1/2$ disclinations. For this configuration, Eq. \eqref{eqn:ShearLine} reduces to
\begin{equation} \label{eqn:2DShearVel}
    \mathbf{v} = 2\gamma y \mathbf{\hat{x}} - 2 m \frac{4 \lambda \gamma a}{3 S_N}\mathbf{\hat{y}}
\end{equation}
where $m = \pm 1/2$ is the topological charge of the disclination. The first term in Eq. \eqref{eqn:2DShearVel} is advection by the flow field, while the second arises from the tendency for the nematogens to flow align in shear flows. Just as with the applied field, the motion in the $\mathbf{\hat{y}}$ direction changes depending on the topological charge of the disclination $m$, assuming disclinations are oriented identically. We note that this result was also predicted for a purely two-dimensional liquid crystal in Ref. \cite{angheluta21}.

\begin{figure}
    \centering
    \includegraphics[width = \columnwidth]{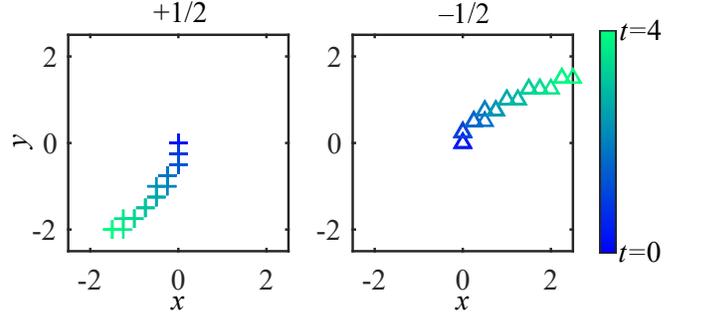}
    \caption{Trajectories of $\pm 1/2$ wedge disclinations in the presence of the shear flow $\mathbf{u} = \gamma y \mathbf{\hat{x}}$ for $\gamma = 2$. As predicted in Eq. \eqref{eqn:2DShearVel}, $\pm$ 1/2 disclinations move in opposite directions depending on their charge. The color indicates the time (in computational units).} 
    \label{fig:ShearDisclinations}
\end{figure}

To test this result, the full Beris-Edwards equations, Eq. \eqref{eqn:BerisEdwards}, for a nematic are solved under an imposed shear flow $\mathbf{u} = \gamma y \mathbf{\hat{x}}$. We compute the effect of flow on $\pm 1/2$ disclinations initially located at the origin. For the computation we set $\Delta t = 0.1$, $\lambda = 1$, and $\gamma = 2$. Figure \ref{fig:ShearDisclinations} shows the trajectories of the disclinations over time. The disclinations have a component of their velocity along either the $\pm \mathbf{\hat{y}}$ directions depending on their charge, which is predicted by Eq. \eqref{eqn:2DShearVel}. Much like the case of the applied field, these results demonstrate that an applied shear flow may be used as a sorting agent for disclinations of opposite charge, while the full behavior, Eq. \eqref{eqn:ShearLine} may be used to predict the motion of disclinations in the more general scenario. 

Note that there is an asymmetry between the trajectories of the $+1/2$ and $-1/2$ disclinations in Fig. \ref{fig:ShearDisclinations}. This can be understood as a higher order effect resulting from the change in director near the core. From the Beris-Edwards equations, Eq. \eqref{eqn:BerisEdwards}, the rate of change of the director will be dependent on $S$ (and $P$). Thus, since $\nabla S \neq 0$ at the core, there will be, over time, a nonzero $\nabla \tphi$. This will then lead to additional motion of the disclination as per the discussion of Sec. \ref{sec:Approximations}. Because this effect is due to the relaxational terms of the dynamics, we expect that if $\gamma \gg 1/\tau$ (that is, the shear flow is much faster than the relaxation rate of the nematic) this effect will be negligible. This is likely the case in experiments involving shear flow \cite{baza20}. In our numerics, however, we have set $\gamma \sim 1/\tau$ due to computational limitations. 

Finally, it is possible to obtain the evolution under shear flow of a disclination loop in a similar manner to the prediction of the external field. Assuming the loop lies in the $xy$ plane, we find
\begin{multline} \label{eqn:shearLoop}
    \mathbf{v} = 2\bm{\hat{\theta}} \times \left(\mathbf{u}\times \bm{\hat{\theta}}\right) - \frac{3}{2R}\bm{\hat{\rho}} \\ + \frac{2\lambda \gamma a}{3 S_N}\biggl[\Bigl( \left(\w\cdot\n_0\right)\left(\mathbf{\hat{u}}\cdot\n_0\right) - \left(\w\cdot \n_1\right)\left(\mathbf{\hat{u}}\cdot\n_1\right)\Bigr)\bm{\hat{\rho}} \\ - \Bigl(\left(\w\cdot\n_0\right)\left(\mathbf{\hat{u}}\cdot\n_1\right) + \left(\w\cdot\n_1\right)\left(\mathbf{\hat{u}}\cdot\n_0\right)\Bigr)\mathbf{\hat{z}}\biggr]
\end{multline}
where we have included the self-annihilation term predicted in Eq. \eqref{eqn:LoopVelocity}. Equation \eqref{eqn:shearLoop} is similar to Eq. \eqref{eqn:ShearLine}, except the tangent vector is $\T = \bm{\hat{\theta}}$ and the directions of motion are $\bm{\hat{\rho}}$ and $\mathbf{\hat{z}}$ as in Eq. \eqref{eqn:HLoop}. Equation \eqref{eqn:shearLoop} predicts many similar features to that of the applied field, except for the addition of advection. For instance, fixing the shear flow directions to $\w = \mathbf{\hat{z}}$ and $\mathbf{\hat{u}} = \mathbf{\hat{y}}$ and setting $\OMEGA = \mathbf{\hat{x}}$ gives similar results to the applied field, and the motion is dependent on the angle between $\n_0$ and the $z$-axis. If $\n_0 = \mathbf{\hat{z}}$, the loop is predicted to have a component of the velocity in the $+\mathbf{\hat{z}}$ direction. On the other hand, if $\n_0 = \mathbf{\hat{y}}$ the loop is predicted to move oppositely. If $\n_0$ is skewed from these two directions, there will be an additional velocity in the $\bm{\hat{\rho}}$ direction. Particularly, if $\n_0 \propto \mathbf{\hat{y}} + \mathbf{\hat{z}}$, there will be an unstable critical radius above which the loop will continue to grow. 

\begin{figure}
    \centering
    \includegraphics[width = \columnwidth]{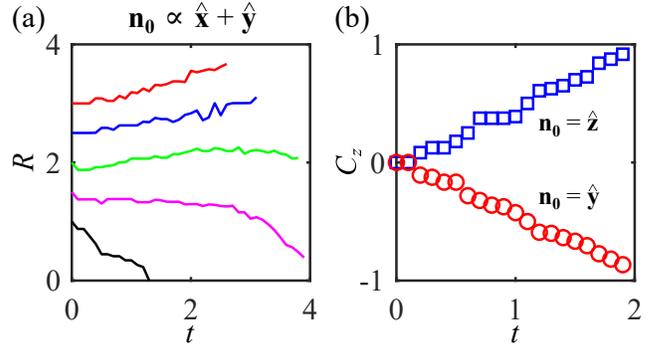}
    \caption{Motion of wedge-twist disclination loops ($\OMEGA = \mathbf{\hat{x}}$) in shear flow $\mathbf{u} = \gamma z \mathbf{\hat{y}}$ for $\gamma = 2$. (a) Loop radius $R$ versus time $t$ for various initial radii with $\n_0 \propto \mathbf{\hat{x}} + \mathbf{\hat{y}}$. The radius of the loop is calculated by finding the average distance between the center of the loop and points in which $S < 0.3S_N$. Above $R \approx 2$ the loop disclinations grow indefinitely. (b) $z$ coordinate of the center of the loop, $C_z$, versus $t$ for the cases $\n_0 = \mathbf{\hat{z}}$ and $\n_0 = \mathbf{\hat{y}}$. For these cases, depending on $\n_0$, the loop has a positive or negative velocity component in the $\mathbf{\hat{z}}$ direction.}
    \label{fig:ShearLoops}
\end{figure}

We have numerically tested these predictions for the cases outlined above for a full three-dimensional disclination loop under shear flow. We perform the same analysis as with the applied field: Fig. \ref{fig:ShearLoops}(a) shows the radius of the loop as a function of time for various initial radii when $\n_0 = (1/\sqrt{2})(\mathbf{\hat{y}} + \mathbf{\hat{z}})$, and Fig. \ref{fig:ShearLoops}(b) shows the $z$ component of the center of the loop disclination, $C_z$, for the cases $\n_0 = \mathbf{\hat{z}}$ and $\n_0 = \mathbf{\hat{y}}$. For the computations, we use the same parameters as indicated above. We see from the figures that the motion of the disclination loops are qualitatively well described by Eq. \eqref{eqn:shearLoop}.

That coupling to shear flow yields similar disclination motion to that of an applied field is a striking, yet intuitive result. For a flow aligning nematic, there is a steady state in which the director everywhere is given by some angle to the flow \cite{vanhorn00,murza18}. Thus, flow has a similar effect on a uniform nematic to an applied field and so it is not surprising that it would have similar effects on disclinations. However, theoretically (and in particular analytically) proving that the effects are similar on disclinations is not simple due to the nonlinear nature of the equations, and the fact that the system is out of equilibrium. This demonstrates the power of the kinematic description of disclinations, which allows analytical calculation even in non-equilibrium systems.

It is understandable that a flow aligning nematic should behave similarly to that under an applied field that also aligns the nematic, yet Eqs. \eqref{eqn:ShearLine} and \eqref{eqn:shearLoop} do not distinguish between the case of flow aligning and flow tumbling. If the tumbling parameter
\begin{equation}
|\lambda| < \frac{6 S_N}{2 S_N + 4}
\end{equation}
the nematic will be in a flow tumbling state, that is, a state in which the director continuously rotates with the flow instead of aligning with it \cite{yeomans16,murza18}. However, Eqs. \eqref{eqn:ShearLine} and \eqref{eqn:shearLoop} predict that reducing the tumbling parameter simply reduces the speed of the disclinations. We have verified that this is indeed the case numerically when the nematic is in the flow tumbling regime, $\lambda = 0.5$. In this case we find $\pm 1/2$ wedge disclinations move similarly to those shown in Fig. \ref{fig:ShearDisclinations}.

One aspect of Eq. \eqref{eqn:shearLoop} which is different from the case of an applied field, Eq. \eqref{eqn:HLoop}, is the advection term. If the disclination loop is out of the plane of the flow direction, then the effect of the advection will be to rotate the loop to lie in this plane. The implications for twist disclinations in particular are interesting, since, if in the flow direction plane, a twist disclination will not be affected by the flow. This may be the mechanism in experiments in which twist disclinations are nucleated and observed at higher rates than wedge-twist disclinations, for both passive nematics in shear flow and active nematics \cite{duclos20,baza20,zhang21}.

\section{Discussion}\label{sec:Discussion}
We have derived a kinematic equation in terms of the nematic tensor order parameter at a disclination core that gives the velocity of three dimensional disclination lines in nematic liquid crystals. In particular, we have generalized previous results involving order parameter singularity tracking methods to a case in which the order parameter is not singular at the defect core. The focus, instead, are zeros in a subspace of the order parameter space. Further, we have made use of the property that this velocity only depends on derivatives of the order parameter at the core to outline an approximation scheme that allows for analytic predictions of the disclination velocity in a number of cases of interest.

The geometric complexity of three dimensional disclination lines has proven a barrier in the study of even the simplest configurations containing multiple disclinations. Here we have presented analytic results which either reinterpret previous results or yield novel predictions for disclination motion. They include equations for the motion of twisted disclinations in two dimensions, recombining and rotating disclination lines in three dimensions, self-annihilating disclination loops with both isotropic and anisotropic elasticity, and disclination lines and loops under externally applied fields and flows. All of these configurations are of current interest and importance in studies of the coarsening and control of passive and active nematics.

The results and methods presented in this work should be of considerable use to experimental and large scale simulation studies in which disclinations play important roles. As demonstrated in Sec. \ref{subsec:ExternalFieldsFlows}, applying fields or flows to systems of disclinations allows the identification of disclination geometry and topology without the explicit recreation of the order parameter. Additionally, the methods may be used to engineer disclination motion which can be further used as particle aggregators and transporters. 

One aspect of disclination evolution that still needs resolution is the evolution of the rotation vector. Because it is not topologically protected, the rotation vector may change as a function of time. This behavior is different from the seemingly similar \cite{kleman89,long21} Burgers vector characterizing dislocations in solids, which remains constant. A detailed study of how this geometric property evolves in time for various configurations will be important as it is intricately connected to the dynamics of disclination lines. Additionally, while we studied circular loops here with constant $\OMEGA$, disclinations in experiments are not perfectly circular and have rotation vectors that vary throughout the loop. Thus, a more general treatment of curvature and torsion of disclination lines as well as varying $\OMEGA$ may lead to a better understanding of their overall dynamics.

Another interesting area that we have not studied here is active nematics. These systems have been heavily studied in two dimensions \cite{marchetti13,doo16,doo18,opathalage19,shankar19,angheluta21} and three-dimensional active nematics are recently receiving more attention \cite{duclos20,binysh20,houston21}. In active nematics, disclinations play a crucial role by driving flows into a mesoscopic chaotic state. Thus, understanding the behavior of disclinations is highly important in this field. However, due to the chaotic nature of the flows, and the complex interaction with the geometric properties of disclinations, very little has been studied analytically regarding systems of many flowing disclinations. The kinematic equation presented here may prove to be a useful tool in understanding individual disclination dynamics in these out of equilibrium systems.

Finally, while we have focused here on the nematic phase, the methods presented in Secs. \ref{sec:LineKinematics} and \ref{sec:Approximations} can be generalized to other system with broken symmetries that support topological defects. The derivation of kinematic equations similar to Eq. \eqref{eqn:Velocity} may yield fruitful analogies between otherwise physically dissimilar systems. For example, an equivalent to Eq. \eqref{eqn:Velocity} was recently derived for dislocations in solids using a phase field approach \cite{skogvoll22}. Topological defects are also important in high energy physics and cosmology \cite{kibble97,pismen99}. Comparison of kinematic velocities in these fields to condensed matter systems might allow for more cross disciplinary experiments. 

\begin{acknowledgments}
We are indebted to Jonathan Selinger for clarifying the scope of the definition of the tensor $\mathbf{D}$. His comments led to a redefinition of its path integral around a defect core from the earlier version given in \cite{schimming22}, which is now detailed in Appendix \ref{append:DensityTensor}. We are also indebted to him and Daniel Beller for many useful comments about this work. This research has been supported by the National Science Foundation under Grant No. DMR-1838977, and by the Minnesota Supercomputing Institute. C.D.S. also gratefully acknowledges support from the U.S. Department of Energy through the LANL/LDRD Program and the Center for Nonlinear Studies for part of this work.

\end{acknowledgments}

\appendix

\section{The $\mathbf{D}$ Tensor}\label{append:DensityTensor}
The tensor quantity $\mathbf{D}$ appearing in the Jacobian transformation of Eq. \eqref{eqn:DensityQ} has a deeper connection to the topological charge of a disclination than simply connecting the disclination densities. Here we expand upon this connection which was first explored in Ref. \cite{schimming22}. The arguments in this appendix are slightly altered from the arguments in \cite{schimming22}, but the mathematical definitions and results remain unchanged. 

We begin this exploration by first recalling the topological charge definition in two dimensions in terms of three different nematic parameters; namely, the angle of the director $\phi$, the director $\n$, and the tensor $\Q$:
\begin{align} \label{eqn:TopCharge2D}
    m &= \frac{1}{2\pi}\oint_C \partial_k \phi \, d\ell_k \nonumber \\
    &= \frac{1}{2\pi}\oint_C \varepsilon_{\mu \nu} \hat{n}_{\mu} \partial_k \hat{n}_{\nu} \, d\ell_k \\
    &= \frac{1}{2\pi S_N^2} \oint_{C^*} \varepsilon_{\mu \nu} Q_{\mu \alpha}\partial_k Q_{\nu \alpha} \nonumber
\end{align}
where $C^*$ represents a closed curve that is restricted to regions in which $S = S_N$ is constant. For all of these cases, a ``topological density'' may be defined by invoking Stokes' theorem. For the case of the singular quantities $\phi$ and $\n$, these densities are Dirac delta-functions located at cores of defects. For the case of $\Q$, the density is a diffuse scalar field with maxima or minima located at defect cores. The scalar field density has been used to identify and track defects in various previous studies of nematics \cite{blow14}.

Generalizing Eq. \eqref{eqn:TopCharge2D} to three dimensional disclination lines is a difficult task due to the added dimension in which the director can lie. In the three-dimensional case, the nematic ground state manifold is equivalent to a hemisphere in which all points on the equator are identified with their polar opposites \cite{alexander12}. A curve through a nematic in real space may be mapped to a curve on the ground state manifold. 

To generalize Eq. \eqref{eqn:TopCharge2D}, recall that the topological charge of a line disclination is always $+1/2$. We thus seek to construct an integral that gives $+1/2$ or zero, modulo $2\pi$. To do this, we construct a locally defined unit vector $\XI$ so that $\XI \cdot \left(\n\times d\n\right)$ gives only the projected length of the curve in the ground state manifold. The quantity $\n\times d\n$ gives the rotation of the director about the axis projected into it \cite{efrati14}. The vector $\XI$ is then chosen so that the rotation about a great circle is measured. The idea is that if one only integrates the contribution of arc length along a great circle, the total integral will either be the length of the great circle from opposite points on the equator (i.e. $\pi$) or zero if the curve does not pass through the equator, since equal contributions will move in opposite directions along the projection.

Explicitly, after mapping the charge measuring curve in real space to the ground state manifold, an arbitrary point on the curve $\n^*$ is fixed. Then, for each point $s$ on the curve
\begin{equation}
    \XI(s) \equiv \frac{\n(s) \times d\n_{||}(s)}{|\n(s) \times d\n_{||}(s)|}
\end{equation}
is computed. Here $d\n_{||}(s)$ is the tangent vector of a great circle defined by the fixed point $\n^*$ and the current point $\n(s)$. Then
\begin{equation} \label{eqn:SigDef}
    \XI(s) \cdot \left(\n(s) \times d \n(s)\right) = \frac{d\n(s) \cdot d\n_{||}(s)}{|\n(s) \times d\n_{||}(s)|} \equiv d\varsigma(s)
\end{equation}
where we have introduced the notation $\varsigma(s)$ to indicate the projected arclength along a great circle.

We claim that
\begin{equation} \label{eqn:3DDiscCharge}
    \oint_C d\varsigma(s) \in \left\{0,\frac{1}{2}\right\}\,\, \text{modulo}\,\, 2\pi.
\end{equation}
To show this, we first assume we are working locally on the unit sphere and define the unit vector
\begin{equation}
    \mathbf{\hat{V}}(s) = a(s) \n^* + b(s)\n(s)
\end{equation}
where $a(s)$ and $b(s)$ are defined such that $\mathbf{\hat{V}}(s)\cdot \n^* = 0$. Given, $\n^*$ and $\n(s)$, this can always be achieved via the Gram-Schmidt procedure. Then the curve
\begin{equation}
    \mathbf{\hat{W}}(t,s) = \cos t \n^* + \sin t \mathbf{\hat{V}}(s)
\end{equation}
parameterizes the great circle on the unit sphere passing through $\n^*$ and $\n(s)$ for any $s$. In particular, we have that $\mathbf{\hat{W}}(t^*(s),s) = \n(s)$ for $t^*(s) = -\arctan\left[1/a(s)\right]$. Then, Eq. \eqref{eqn:SigDef} may be written as
\begin{equation}
    d\varsigma(s) = \left.\frac{d\n/ds \cdot d\mathbf{\hat{W}}/dt}{|\n \times d\mathbf{\hat{W}}/dt|}\right|_{t = t^*(s)} ds.
\end{equation}
Substituting this into Eq. \eqref{eqn:3DDiscCharge} and simplifying yields
\begin{equation} \label{eqn:IntOfSigma}
    \oint_C d\varsigma(s) = \n^* \cdot \oint_C \frac{d\n}{|\n^* \times \n|}. 
\end{equation}

If the curve $C$ in Eq. \eqref{eqn:IntOfSigma} does not cross the equator of the unit sphere the integral is zero since the curve starts and ends at the same point. However, if the curve does cross the equator, then $\pi$ must be added to the contour integral for each time the equator is crossed. $\pi$ must added because this is the length of the arc of a great circle connecting the two points on the equator. A subtle point here is that if the equator is passed an even number of times the measured configuration in three dimensions is topologically equivalent to a configuration with no disclinations. This is not represented by our measure since we are representing the ground state manifold (namely, the real projective plane) with vectors and so we must manually take the resulting calculation modulo $2 \pi$.

We may then write the equation for the charge of a disclination in the following ways:
\begin{align} \label{eqn:3DQCharge}
    m &= \oint_C \hat{\Xi}_{\gamma} \varepsilon_{\gamma \mu \nu} \hat{n}_{\mu} \partial_k \hat{n}_{\nu} \, d \ell_{k} \,\, \text{modulo}\, \, 2 \pi \\
    &= \frac{1}{S_N^2} \oint_{C^*} \hat{\Xi}_{\gamma} \varepsilon_{\gamma \mu \nu} Q_{\mu \alpha} \partial_k Q_{\nu \alpha}\, d \ell_{k} \,\, \text{modulo} \, \, 2 \pi. \nonumber
\end{align}
As the measuring curve $C$ is taken to be smaller and smaller, the resulting curve in the ground state manifold approaches a great circle, since near the disclination $\n$ is given by Eq. \eqref{eqn:nCore}. In this case $\XI \to \OMEGA$. Thus we identify $\OMEGA$ as a geometric property of the disclination core. 

Eq. \eqref{eqn:3DQCharge} may be used to detect line disclinations, however, in practice it is complicated to compute $\XI$ for various curves. Additionally, many curves must be constructed to measure the full extent of a disclination line. Therefore, just as is done in two dimensions, we construct a topological density by applying Stokes' theorem to Eq. \eqref{eqn:3DQCharge} and taking the integrand of the resulting integral. Near the core of a disclination and from arguments similar to the derivation of Eq. \eqref{eqn:DensityQ}, we are left with a single term:
\begin{equation}
    \hat{\Xi}_{\gamma} \varepsilon_{\gamma \mu \nu} \varepsilon_{i k \ell} \partial_k Q_{\mu \alpha} \partial_{\ell} Q_{\nu \alpha} \equiv \XI \cdot \mathbf{D}.
\end{equation}
This serves as a definition for the tensor field $\mathbf{D}$ that appears in Eq. \eqref{eqn:DensityQ}. We are thus able to derive it in two ways: through direct computation of the topological charge, or through the Jacobian transformation from real space to the subspace of order parameter space that vanishes at disclination locations as in Sec. \ref{sec:LineKinematics}.

We note that another term may be nonzero in the topological density defined by applying Stokes' theorem to Eq. \eqref{eqn:3DQCharge}. This term is $\varepsilon_{i k \ell} \partial_k \hat{\Xi}_{\gamma} \varepsilon_{\gamma \mu \nu} Q_{\mu \alpha} \partial_{\ell} Q_{\nu \alpha}$. Away from defects, where it is possible $\XI \neq \OMEGA$, this term may be nonzero in order to cancel a nonzero $\mathbf{D}$ that arises from double-splay or double-twist configurations \cite{selinger19,long21b}. Since a double-splay or double-twist configuration is not a disclination, the integral in Eq. \eqref{eqn:3DQCharge} must give zero and so these terms must cancel. That $\mathbf{D}$ is nonzero for these configurations is an interesting result, and more work is needed to fully understand this, though we speculate that it is likely due to the fact that $\mathbf{D}$ is related to the Jacobian transformation, relating real space areas to areas of subspaces of order parameter space. Note that patches of double-splay and double-twist configurations can be mapped to patches of ``area'' on the unit sphere.

For line disclinations, there is no double-splay or double-twist configuration. Using the linear core approximation near the core, Eq. \eqref{eqn:QApprox}, the $\mathbf{D}$ tensor is computed to decompose as 
\begin{equation}
    \mathbf{D}(\mathbf{r}) = |\mathbf{D}|\left(\OMEGA \otimes \T\right)
\end{equation}
and hence can be used as a disclination geometry identifier. As has already been investigated \cite{schimming22}, the decomposition also holds away from disclinations and $|\mathbf{D}| = 0$ as long as there is no double-splay or double-twist configuration.

\section{Disclination Velocity Derivation}\label{append:Velocity}
Here we derive the disclination velocity presented in Sec. \ref{subsec:Velocity}. This derivation follows previous results for line defects from Halperin and Mazenko. The difference here is that the order parameter is a tensor and it does not go to zero at the disclination core. Many of the manipulations are similar to Refs. \cite{liu92,mazenko97,mazenko99}.

We begin by showing that the $\mathbf{D}$ tensor is a conserved quantity. Namely,
\begin{equation}
    \partial_t D_{\gamma i} = 2 \partial_k\left[\varepsilon_{\gamma \mu \nu}\varepsilon_{i k \ell}\partial_t Q_{\mu \alpha} \partial_{\ell} Q_{\nu \alpha}\right] \equiv 2 \partial_k J_{\gamma i k}
\end{equation}
where $\mathbf{J}$ represents a topological current and we recall that $\Q$ is a regular quantity so that $\varepsilon_{ik\ell}\partial_k \partial_{\ell} Q_{\mu \alpha} = 0$. Using properties of Dirac delta functions we may then write
\begin{align}
    \partial_t\mathbf{D} \delta\left[\Q_{\perp}\right] &= 2\nabla \cdot \mathbf{J} \delta\left[\Q_{\perp}\right] \\
    \Leftrightarrow \mathbf{D} \partial_t \delta\left[\Q_{\perp}\right] &= 2 \mathbf{J} \cdot \nabla \delta\left[\Q_{\perp}\right]. \nonumber
\end{align}
Using this identity, we write a continuity equation for the topological charge density by taking a time derivative of Eq. \eqref{eqn:DensityQ}:
\begin{equation} \label{eqn:Qdtrho}
    \begin{gathered}
    \partial_t \rho_i = \partial_t\delta\left[\Q_{\perp}\right]\hat{\Omega}_{\gamma} D_{\gamma i} + \delta\left[\Q_{\perp}\right] \partial_t \hat{\Omega}_{\gamma} D_{\gamma i} + \delta\left[\Q_{\perp}\right] \hat{\Omega}_{\gamma}\partial_t D_{\gamma i} \\
    = 2\partial_k\delta\left[\Q_{\perp}\right]\hat{\Omega}J_{\gamma i k} + 2\delta\left[\Q_{\perp}\right] \hat{\Omega}_{\gamma i k} \partial_k J_{\gamma i k} \\
    = 2 \partial_k\left(\delta\left[\Q_{\perp}\right] \hat{\Omega}_{\gamma} J_{\gamma i k}\right)
    \end{gathered}
\end{equation}
where we have used that $\partial_t \OMEGA \cdot \mathbf{D} = 0$ since $\OMEGA$ is a unit vector proportional to the first vector component of $\mathbf{D}$ (see Appendix \ref{append:DensityTensor}).

Eq. \eqref{eqn:Qdtrho} connects the current of the $\mathbf{D}$ tensor with the topological charge current, specifically at the location of the disclination core. We now compute the topological charge current by taking a time derivative of Eq. \eqref{eqn:DiscDens}:
\begin{equation}
    \partial_t \rho_i = \frac{1}{2}\int \frac{dv_i}{ds}\delta\left[\mathbf{r}-\mathbf{R}\right] \, ds + \frac{1}{2}\int \hat{T}_i \partial_t \delta\left[\mathbf{r}-\mathbf{R}\right]\, ds
\end{equation}
where we have used $d\mathbf{R}/ds = \T$ and $\partial_t\mathbf{R} = \mathbf{v}$. Manipulating this expression by integrating by parts and changing integration variables from $s$ to $\mathbf{R}$ gives
\begin{align} \label{eqn:rhodtrho}
    \partial_t\rho &= \frac{1}{2}\partial_k \left( v_i\int\delta\left[\mathbf{r} - \mathbf{R}\right] \, dR_k - v_k \int\delta\left[\mathbf{r} - \mathbf{R}\right] \, dR_i\right) \\
    &= \partial_k\left(v_i\rho_k - v_k \rho_i\right).\nonumber
\end{align}
The expression for the topological charge current in the second line is antisymmetric, so the velocity will be perpendicular to $\bm{\rho}$, as it should.

We then compare Eqs. \eqref{eqn:Qdtrho} and \eqref{eqn:rhodtrho} and substitute Eq. \eqref{eqn:DensityQ} for $\bm{\rho}$ to give (up to the curl of a vector field)
\begin{equation} \label{eqn:JEJ}
    2 \hat{\Omega}_{\tau} J_{\tau i k}\delta\left[\Q_{\perp}\right] = \hat{\Omega}_{\gamma}\left(v_i D_{\gamma k} - v_k D_{\gamma i}\right)\delta\left[\Q_{\perp}\right].
\end{equation}
We finally write $J_{\gamma i k} = \varepsilon_{i k \ell} g_{\gamma \ell}$ to define the tensor $\mathbf{g}$ in Eq. \eqref{eqn:Velocity}. Rearranging Eq. \eqref{eqn:JEJ} gives the velocity in Eq. \eqref{eqn:Velocity}:
\begin{equation}
    \mathbf{v}(s) = 2 \left.\frac{\T \times \left(\OMEGA \cdot \mathbf{g}\right)}{|\mathbf{D}|}\right|_{\mathbf{r} = \mathbf{R}(s)}.
\end{equation}

\section{Analytic Computation of Disclination Velocity}\label{append:Analytic}
In this appendix, we complete the details of the computation to go from the approximation of $\Q$ at the core, Eq. \eqref{eqn:FullQApprox}, to the disclination velocity in the presence of inhomogeneous local rotation, Eq. \eqref{eqn:RotateMotion}. The details of this calculation can be used to reproduce the other calculations presented in Sec. \ref{sec:Results}.

To use Eq. \eqref{eqn:Velocity} we must compute $\partial_t Q_{\mu \nu} = \nabla^2 Q_{\mu \nu}$ and $\nabla Q_{\mu \nu}$ at the disclination core ($x = y = 0$). We first compute $\nabla\Q$ in the vicinity of the core,
\begin{multline}
    \nabla Q_{\mu\nu} = S_N\bigg[\frac{\mathbf{\hat{x}}}{2a}\left(\tilde{n}_{0\mu}\tilde{n}_{0\nu} - \tilde{n}_{1\mu}\tilde{n}_{1\nu}\right) \\+  \frac{\mathbf{\hat{y}}}{2a}\left(\tilde{n}_{0\mu}\tilde{n}_{1\nu} + \tilde{n}_{1\mu}\tilde{n}_{0\nu}\right)  \\
     + \frac{x}{2a}\nabla\tphi \left(p_{0\mu}\tilde{n}_{0\nu} + \tilde{n}_{0\mu}p_{0\nu} - p_{1\mu}\tilde{n}_{1\nu} - \tilde{n}_{1\mu}p_{1\nu}\right)  \\
     + \frac{y}{2a}\nabla\tphi\left(p_{0\mu}\tilde{n}_{1\nu} + \tilde{n}_{0\mu}p_{1\nu} + p_{1\mu}\tilde{n}_{0\nu} + \tilde{n}_{1\mu}p_{0\nu}\right)\bigg]
\end{multline}
where we recall $\mathbf{\tilde{n}}_k = \n_k + \tphi(\q \times \n_k)$ and we define $\mathbf{p}_k \equiv \q \times \n_k$. Then we have, at the disclination core,
\begin{multline}
    \left.\nabla^2 Q_{\mu\alpha} \right|_{x=y=0} \\ = S_N \bigg[ \frac{\partial_x \tphi}{a}\left(p_{0\mu}\tilde{n}_{0\alpha} + \tilde{n}_{0\mu}p_{0\alpha} - p_{1\mu}\tilde{n}_{1\alpha} - \tilde{n}_{1\mu}p_{1\alpha}\right) \\
    + \frac{\partial_y \tphi}{a}\left(p_{0\mu}\tilde{n}_{1\alpha} + \tilde{n}_{0\mu}p_{1\alpha} + p_{1\mu}\tilde{n}_{0\alpha} + \tilde{n}_{1\mu}p_{0\alpha}\right)\bigg] \\
    \left. \nabla Q_{\nu \alpha} \right|_{x=y=0} = S_N\bigg[ \frac{\mathbf{\hat{x}}}{2a}\left(\tilde{n}_{0\nu}\tilde{n}_{0\alpha} - \tilde{n}_{1\nu}\tilde{n}_{1\alpha}\right) \\
    + \frac{\mathbf{\hat{y}}}{2a}\left(\tilde{n}_{0\nu}\tilde{n}_{1\alpha} + \tilde{n}_{1\nu}\tilde{n}_{0\alpha}\right)\bigg].
\end{multline}
To compute $\OMEGA\cdot \mathbf{g} = \hat{\Omega}_{\gamma}\varepsilon_{\gamma\mu\nu}\nabla^2 Q_{\mu \alpha} \nabla Q_{\nu \alpha}$ (note this is a vector quantity), we make use of the following relations between the $\mathbf{\tilde{n}}_k$ and $\mathbf{p}_k$ to $O(\tphi)$:
\begin{equation}
   \begin{gathered}
    \mathbf{\tilde{n}}_0 \cdot \mathbf{\tilde{n}}_1 = 0, \quad |\mathbf{\tilde{n}}_0|^2 = |\mathbf{\tilde{n}}_1|^2 = 1, \quad \mathbf{p}_k, \cdot \mathbf{\tilde{n}}_k = \tphi|\q \times \n_k|^2,\\
    \mathbf{p}_0 \cdot \mathbf{\tilde{n}}_1 = \q\cdot\OMEGA - \tphi\left(\q\cdot\n_0\right)\left(\q\cdot\n_1\right), \\
    \mathbf{p}_1 \cdot \mathbf{\tilde{n}}_0 = -\q\cdot\OMEGA - \tphi\left(\q\cdot\n_0\right)\left(\q\cdot\n_1\right), \\
    \mathbf{\tilde{n}}_0 \times \mathbf{\tilde{n}}_1 = \OMEGA, \quad \mathbf{p}_k \times \mathbf{\tilde{n}}_k = -\q + \n_k\left(\q\cdot\n_k\right) \\
    \mathbf{p}_0 \times \mathbf{\tilde{n}}_1 = \n_0\left(\q\cdot\n_1\right) - \tphi\mathbf{p}_1\left(\q\cdot\n_0\right), \\
    \mathbf{p}_1 \times \mathbf{\tilde{n}}_0 = \n_1\left(\q\cdot\n_0\right) + \tphi\mathbf{p}_1\left(\q\cdot\n_1\right).
    \end{gathered}
\end{equation}
We then arrive at 
\begin{equation}
    \OMEGA \cdot \mathbf{g} = \frac{2 S_N^2 \left(\q\cdot\OMEGA\right)}{a^2}\mathbf{\hat{z}}\times\left(\mathbf{\hat{z}}\times \nabla\tphi\right)
\end{equation}
which can be substituted into Eq. \eqref{eqn:Velocity} to get the final velocity of the disclination, Eq. \eqref{eqn:RotateMotion}. The last piece that is needed is $|\mathbf{D}|$ at the core which can be computed from $\nabla \Q$ above to give $|\mathbf{D}(\mathbf{0})| = S_N^2 / a^2$. 

As mentioned, similar manipulations apply to all calculations presented in Sec. \ref{sec:Results}. Note that for configurations involving disclination loops we use cylindrical coordinates to represent real space, and hence the derivatives must be computed using corresponding formulae. In the computations, this leads to a clear distinction between terms in the velocity that arise from disclination curvature---terms that do not include $\nabla\tphi$---and terms that arise from interaction with the rest of the loop. It is likely possible to extend these computations to disclinations with arbitrary curvature and torsion by assigning a Frenet frame to the disclination; however, such calculations are beyond the scope of this work.

\section{Computational Model Details}\label{append:CompDetails}
Here we give details regarding the computational model used in Sec. \ref{sec:Results}. Specifically we will focus on the implementation of Eq. \eqref{eqn:BallMajumdar}. 

To numerically implement the bulk free energy, we must write it in terms of the order parameter $\Q$. To accomplish this, given a value of $\Q$ (or a field $\Q(\mathbf{r})$), $\Delta s$ is maximized with respect to the probability distribution $p(\bm{\hat{\xi}})$ subject to the constraint that $\Q$ is given by Eq. \eqref{eqn:QDef}. To impose the constraint, a tensor Lagrange multiplier $\bm{\Lambda}$ is introduced so that the optimal probability distribution is given by
\begin{equation} \label{eqn:ProbDist}
    p^*(\bm{\hat{\xi}}) = \frac{e^{\left[\bm{\hat{\xi}}^T \bm{\Lambda}\bm{\hat{\xi}}\right]}}{Z}, \quad Z = \int_{{\cal S}^2} e^{\left[\bm{\hat{\xi}}^T \bm{\Lambda}\bm{\hat{\xi}}\right]}\,d\Sigma(\bm{\hat{\xi}})
\end{equation}
where $Z$ is interpreted as a single particle partition function.

The Lagrange multiplier $\bm{\Lambda}$ can be related to the order parameter $\Q$ via the self-consistency equation
\begin{equation} \label{eqn:SelfConsist}
    \frac{\partial \ln Z}{\partial \bm{\Lambda}} = \Q + \frac{1}{3}\mathbf{I}.
\end{equation}
Thus $\bm{\Lambda}$ may be regarded as a function of $\Q$ if Eq. \eqref{eqn:SelfConsist} is inverted. In general, however, the self-consistency relation cannot be analytically inverted, and so it must be done numerically. Substituting Eq. \eqref{eqn:ProbDist} into the entropy density, Eq. \eqref{eqn:BallMajumdar}, leads to a bulk free energy that is dependent only on $\Q$:
\begin{equation} \label{eqn:BulkFreeE}
    f_B(\Q) = -\kappa \text{Tr}\left[\Q^2\right] + n k_B T \bm{\Lambda}:\left(\Q + \frac{1}{3}\mathbf{I}\right) - \ln Z + \ln 4\pi
\end{equation}
where $\bm{\Lambda}$ is treated as a function of $\Q$.

$\delta F/\delta \Q$ can then be calculated and the dynamics $\partial_t \Q = -\delta F/\delta \Q$ are computed using the finite element method described in the main text. One difficult aspect of this model is the computation of $\bm{\Lambda}$ (and $\partial \bm{\Lambda}/\partial \Q$ which is required for the algorithm) given a value of $\Q$. Because $\Q$ varies throughout the system, $\bm{\Lambda}$ must be computed at every node of the mesh for every time step. Fortunately, the self-consistency equation is local, and so we may parallelize the computation of $\bm{\Lambda}$ across the mesh. Specific details of the numerical method are given in Ref. \cite{schimming21}.

As mentioned in the text, the primary advantage of this bulk free energy over the typical Landau-de Gennes free energy is that the free energy will remain bounded for elastic free energies cubic in $\Q$, which is required to model differences between splay and bend. For all of the examples that we present in which splay-bend degeneracy is assumed (i.e. $L_3 = 0$), a Landau-de Gennes bulk free energy should reproduce the results. 

\bibliography{LC}

\end{document}